\documentclass{article}

    \PassOptionsToPackage{numbers, compress}{natbib}

    \usepackage[preprint]{neurips_2024}

\usepackage[utf8]{inputenc} 
\usepackage[T1]{fontenc}    
\usepackage{hyperref}       
\usepackage{url}            
\usepackage{booktabs}       
\usepackage{amsfonts}       
\usepackage{nicefrac}       
\usepackage{microtype}      
\usepackage{xcolor}         
\usepackage{graphicx}       
\usepackage{comment}
\usepackage{wrapfig}
\usepackage{svg}
\usepackage{amsmath} 
\usepackage{subcaption}
\usepackage{pgfplots}

\title{EngineBench: \\ Flow Reconstruction in the Transparent Combustion Chamber III Optical Engine}

\author{%
Samuel J. Baker\footnotemark{}$^{\;,1}$ \quad Michael A. Hobley$^{1}$ \quad Isabel Scherl$^2$ \\ \textbf{XiaoHang Fang}$^{1,3}$ \quad \textbf{Felix C. P. Leach}$^1$ \quad
\textbf{Martin H. Davy}$^1$ \\
$^1$University of Oxford \quad $^2$California Institute of Technology \quad $^3$University of Calgary
}

\begin{document}

\maketitle

\begin{abstract}
We present EngineBench, the first machine learning (ML) oriented database to use high quality experimental data for the study of turbulent flows inside combustion machinery.
Prior datasets for ML in fluid mechanics are synthetic or use overly simplistic geometries.
EngineBench is comprised of real-world particle image velocimetry (PIV) data that captures the turbulent airflow patterns in a specially-designed optical engine.
However, in PIV data from internal flows, such as from engines, it is often challenging to achieve a full field of view and large occlusions can be present.
In order to design optimal combustion systems, insight into the turbulent flows in these obscured areas is needed, which can be provided via inpainting models. 
Here we propose a novel inpainting task using \textit{random edge gaps},  a technique that emphasises realism by introducing occlusions at random sizes and orientations at the edges of the PIV images.
We test five ML methods on random edge gaps using pixel-wise, vector-based, and multi-scale performance metrics.
We find that UNet-based models are more accurate than the industry-norm non-parametric approach and the context encoder at this task on both small and large gap sizes.
The dataset and inpainting task presented in this paper support the development of more general-purpose pre-trained ML models for engine design problems.
The method comparisons allow for more informed selection of ML models  for problems in experimental flow diagnostics.
All data and code are publicly available at \texttt{https://eng.ox.ac.uk/tpsrg/research/enginebench/}.

\end{abstract}

\section{Introduction}
\begin{wrapfigure}{R}{0.45\textwidth}
\begin{center}
\includegraphics[width=0.40\columnwidth]{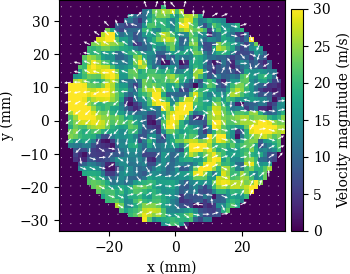}
\end{center}
    \caption{Example PIV image, showing a circular field of view. At each pixel the arrows show the direction of the turbulent flow, and the colourmap shows the velocity magnitude.}
    \label{fig:piv_eg_img}
\end{wrapfigure}

\footnotetext{samuel.baker@eng.ox.ac.uk}
Highly turbulent, internal flows 
underpin crucial technology in a range of sectors from transport and power generation to chemical processing and biofluids. Defined as flows that are bounded by ducts or channels, internal flows are needed in situations where the direction and supply of a fluid needs to be controlled \cite{greitzer2007internal}.
In particular, in the automotive, marine and aerospace industries, these flows are used to power propulsion systems that are essential to the development of highly efficient and low-carbon transport solutions \cite{senecal2021racing,liu2017review}. Experimentally, these flows are often characterized using velocity measurements from particle image velocimetry (PIV) \cite{adrian2011particle}. The PIV method generates velocity vectors at discrete points in the flow, so spatially-dependent flow behaviour can be easily observed and quantified. 
An example of a post-processed image from PIV can be seen in Figure \ref{fig:piv_eg_img}.
Conducting PIV for internal flows presents a unique set of challenges, as it is often difficult to achieve a full field of view. Gaps in the data arise due to shadowing (occlusions due to walls or other components), laser alignment issues, irregular seeding density of the tracer particles, background reflections and light scatter, and strong out-of-plane motion for 2D measurements \cite{scherl2020robust,van2007measurement}. 

Large gaps in PIV data pose a significant barrier to the development of propulsion systems and turbomachinery, as gappy data limit flow field analysis and post-processing, sometimes necessitating experimental reruns and leading to a waste of resources 
\cite{murray2007application}. Further problems arise as design work becomes increasingly digitalised. Simulations of physical equipment produce clean images without gaps, but often require comparison to or assimilation with potentially gappy experimental data for validation purposes \cite{chandramouli20204d,anderson1995computational}. The pursuit of accurate digital twins and simulations of complex geometries has the potential for monumental savings in the time and costs associated with mechanical engineering design, by lessening the need to build and test prototypes \cite{brunton2021data,argyropoulos2015recent}. Developing accurate methods for reconstructing full flow fields from gappy PIV data is therefore a topic of significant interest in the study of industrially-relevant flows and the development of related technologies.

\paragraph{Main contributions}
This work introduces the newly collated EngineBench database, a collection of PIV datasets specially designed for ML research in propulsion systems. EngineBench is currently comprised of data from different experiments previously run on the transparent combustion chamber (TCC-III) optical engine by General Motors and the University of Michigan \cite{schiffmann2016tcc}. The main contributions of this work are listed below.
\begin{itemize}
    \item We present a new collection of previously published datasets, with built-in modularity for ease of use in training ML models (see Section \ref{sec:data}).
    \item We establish the first inpainting benchmark that uses experimental data on an industrially-relevant, turbulence-driven power unit.
    \item We introduce random edge gaps, a novel pipeline for simulating realistic occlusions, which is more representative of  real PIV data from internal flows.
\end{itemize}

\section{Related work}
\label{sec:litrev}
\paragraph{Flow physics datasets}
Several large flow physics datasets exist, including the Johns Hopkins Turbulence Database~\cite{li2008public}, BLASTNet 2.0 database \cite{chung2024turbulence}, and the turbulence data from McConkey \textit{et al.} \cite{mcconkey2021curated}. However, each of these datasets represent idealized canonical flows over small domain sizes, which do not reflect the complex geometries and operating environments associated with physical machinery. Other databases address more practical geometries and domain sizes, such as the AirfRANS dataset for airfoil shape optimisation \cite{bonnet2022airfrans}, and the Cambridge-Sandia burner for a variety of swirling stratified flows \cite{zhou2013flow}. Regarding internal flows, PIV datasets have been published by the Engine Combustion Network (ECN) \cite{meijer2012engine} and the General Motors University of Michigan Automotive Cooperative Research Laboratory \cite{Schiffmann2018}. Data collected by the latter for the TCC-III combustion chamber are currently used in EngineBench, as the TCC-III setup was specifically designed to challenge the predictive capabilities of computational fluid mechanics (CFD) simulations, with the geometry promoting extremely complex fluidic motion via strong turbulence and high cycle-to-cycle variations~\cite{schiffmann2016tcc}. In a development philosophy similar to that of CFD models, it is expected that ML models can be made more generalisable by being trained on highly challenging datasets \cite{ko2021methodology}. These characteristics of the TCC data also proved valuable in the development of non-parametric dimensionality reduction approaches in the early 2010s \cite{chen2012use,chen2013practical}. 

\paragraph{Inpainting for turbulent flows: non-parametric}
The development of numerical methods that can fill gaps in spatio-temporal turbulent flow data has a history spanning several decades. Of particular note is the family of methods stemming from what came to be known as the gappy proper orthogonal decomposition (GPOD), introduced by Everson and Sirovich in 1995 \cite{everson1995karhunen}. These methods employ the POD (akin to the principal component analysis) to identify dominant flow structures in a dataset, which are used to inform the velocity predictions inside the gaps. The predictions are updated iteratively as the number of POD modes (principal components) considered for the reconstruction is incremented. More details on the method can be found in Appendix \ref{app:gpod}. GPOD became an industry-norm in the field of turbulent flow diagnostics, and received several updates and improvements \cite{gunes2006gappy,raben2012adaptive,saini2016development,nekkanti2023gappy}. Part of the reason for GPOD's popularity in the fluid mechanics community is due to the focus that POD methods place on dominant low-rank features, which may be analogous to the concept of coherent structures in turbulent flows \cite{murata2020nonlinear,baker2024extracting,taira2017modal}. Therefore, results from GPOD retain a degree of physical explainability. In addition, as fluid flow data are often negatively affected by noise, outliers, and potentially less relevant small-scale turbulent structures \cite{epps2019singular,roudnitzky2006proper}, high levels of reconstruction accuracy can be achieved by mainly focusing on these low-rank structures \cite{scherl2020robust,saini2016development}. 

\begin{wrapfigure}{R}{0.45\textwidth}
\begin{center}
\includegraphics[width=0.40\columnwidth]{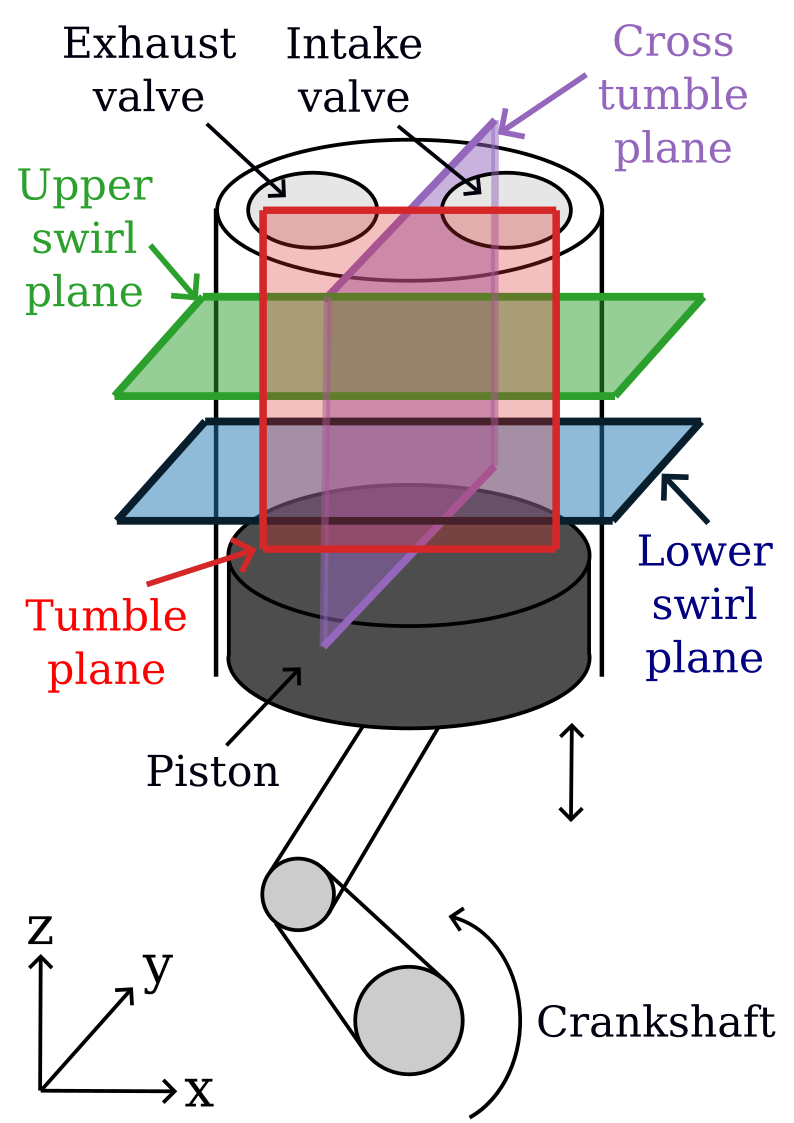}
\end{center}
    \caption{Schematic showing the TCC-III and associated PIV measurement planes. 
    }
    \label{fig:schematic}
\end{wrapfigure}

\paragraph{Inpainting for turbulent flows: parametric}
Deep learning methods are also gaining traction for inpainting tasks in fluid mechanics \cite{vinuesa2023transformative,brunton2020machine}. Autoencoder neural networks have been widely used in turbulent flow applications, partly because the dimensionality reduction through the latent space maintains the focus on dominant flow structures~\cite{dubois2022machine,kumar2022state}. Convolutional neural networks (CNNs) are also widely used due to their ability to utilise the local spatial relationships inherent in turbulent flow data on grids~\cite{nguyen2024climatelearn,jin2018prediction,morimoto2021experimental}. In particular, the UNet architecture~\cite{ronneberger2015u} has demonstrated success in a variety of tasks including flow field prediction and super resolution~\cite{kaltenborn2023climateset,bao2023deep,zhang2023improved,deng2023temporal,ashkboos2022ens}. However, UNets are thought to exhibit some difficulty in capturing the dependency relationships of global features \cite{wang2018non}, due to the relatively slow expansion of the receptive field through the convolutional layers. 
Higher levels of performance have been reported by combining UNets with transformer modules for turbulent flow field prediction and reconstruction from limited measurements~\cite{kang2023new,jiang2023transcfd,xu2023estimation}, due to an enhanced ability to more accurately capturing the multi-scale relationships present in turbulent flows. 

Regarding alternative architectures, generative adversarial networks (GANs) have been used successfully in preserving multi-scale statistics of turbulent flows for super-resolution \cite{guemes2022super,kim2021unsupervised} and inpainting \cite{li2023multi}. Physics-informed neural networks (PINNs) \cite{raissi2019physics} also show promise for creating more generalizable ML models, especially for laminar and fully 3D flows \cite{krishnapriyan2021characterizing,eivazi2022physics,wang2020towards}.
However, the suitability of PINNs for the present 2D PIV setup is currently unclear, as the only data available are two velocity components along a 2D slice of a 3D system, which stretches the validity of the conservation equations. In addition, the spatially correlated noise introduced by the cross-correlation algorithm in the PIV process has been shown to significantly degrade the results from a PINN \cite{wang2022dense}. Other developments such as graph neural networks \cite{kipf2016semi,pfaff2020learning,gao2019graph} and neural operators~\cite{lu2019deeponet,kovachki2023neural,li2020fourier} have demonstrated success in scientific ML applications, but have not seen significant use in inpainting to date \cite{elharrouss2020image,jam2021comprehensive}. 

\paragraph{Gap handling}
In the literature on inpainting for turbulent flows, there is a noticeable lack of discussion on how artificial gaps should be created for training and testing ML models. Common methods of adding gaps to clean data include random noise \cite{venturi2004gappy,wang2015proper}, clustered dropouts \cite{raben2012adaptive,saini2016development}, and block gaps \cite{nekkanti2023gappy,zhang2023improved,luo2023reconstruction}. Rectifying random noise and clustered dropouts is often an easier task for ML models due to the large amount of spatially local information that remains available \cite{buzzicotti2021reconstruction}. Conversely, due to the larger gap sizes, block gaps can be more challenging to handle. However, studies to date have only considered blocks of standard shapes and fixed orientations, which can be unrealistic and of limited use in practical scenarios where complex geometries can obscure the field of view in any number of ways. 

In this work, we introduce \textit{edge gaps}, which simulate a commonly encountered real-world scenario: a reduced field of view due to a restrictive pipe orifice or window \cite{rabault2016study,zha2015characterization}. 

\section{Dataset}
\label{sec:data}
\paragraph{Experimental apparatus}
EngineBench comprises 48 GB of PIV data from the General Motors University of Michigan Automotive Cooperative Research Laboratory TCC-III combustion chamber. The TCC-III is a single-cylinder optical research engine, with a single intake valve, an exhaust valve, and a piston that can move up and down the combustion cylinder in the $z$ direction, as shown in Figure \ref{fig:schematic}. The $z$ position of the piston inside the cylinder is determined by the angle of the rotating crankshaft, so images taken at different piston positions within an engine cycle are referred to as different phases measured in crank angle degrees (CAD). The operation of the apparatus is further explained and additional experimental details are given in Appendix \ref{app:exp}.

A wealth of experience in engine flow PIV systems accumulated since the introduction of TCC-0 in the 1990s \cite{reuss1995particle,reuss2002particle,abraham2013high} was leveraged to make the TCC-III results extremely repeatable across the six month experimental campaign during which the data were gathered. Overall, Schiffmann \textit{et al.} \cite{schiffmann2016tcc} achieved a test-to-test repeatability that exceeded the experimental uncertainty of the measurements, with an average velocity error of 1.5$-$8\% between the mid-intake and mid-compression strokes (90$-$270 CAD) depending on the PIV measurement plane. Data are available for four PIV planes, namely the tumble, cross-tumble, upper-swirl, and lower-swirl planes. These planes are shown in the schematic in Figure~\ref{fig:schematic}. A variety of physical phenomena can be observed on these planes, such as tumble vortex motion, intake jet dynamics, and strong cycle-to-cycle variations. 
The demonstration of complex and multi-physics features such as high turbulence, compressibility, heat transfer, combustion, and moving boundaries
mean that engine flows have often served as useful and challenging test cases for numerical methods, with broad applicability to a range of industrial flows \cite{fogleman2004application,yang2019applicability,zhao2023applications,reitz1995development}.

\paragraph{Format and hosting}
PIV data files are typically not readily formatted for ML model training; they are distributed in a variety of directory structures and file types including \texttt{.txt}, \texttt{.csv} and \texttt{.mat}. Similarly to the BubbleML dataset \cite{hassan2024bubbleml}, we combine the individual PIV files into data matrices and store each matrix as a binary HDF5 file for several reasons, listed below. 
\begin{itemize}
    \item Hierarchical, file-system-like structure for built-in modularity. Simplifies chunking of the data so that train/validation/test splits can be separated by specific phase angles or test points, as employed in this benchmark.
    \item Binary format for efficient data storage and fast I/O speed.
    \item Partial I/O operations for lazy loading and data inspection.
    \item Parallel I/O operations.
    \item Direct synergy with functions that expect matrix inputs, common in fluid mechanics codes.
\end{itemize}

A diagram illustrating the structure of each HDF5 file is given in Appendix \ref{app:h5}. The datasets are publicly hosted on Kaggle, 
and are accompanied by tutorial notebooks to demonstrate how the data can be interacted with. The website 
\texttt{https://eng.ox.ac.uk/tpsrg/research/enginebench}
contains all the necessary links for accessing the data and code.

\paragraph{Licensing}
The study here used publicly available TCC engine data, which was created with funding by General Motors through the General Motors University of Michigan Automotive Cooperative Research Laboratory, Engine Systems Division. The data were released under the Creative Commons Attribution 4.0 International License.

\paragraph{Subset}
In this benchmarking study, a subset of the full database named \texttt{EngineBench\_LSP\_small} is used to simulate a common situation whereby a practitioner has access to a relatively limited amount of data from a single geometry. Furthermore, we consider data solely from one PIV plane and test point. This emulates a scenario whereby moving components obstruct the field of view at some phase angles but not others, allowing an ML model to rectify a gappy phase angle using information from other clean phases from the same experiment. The HDF5 file structure is leveraged to enable training on different permutations of phase angles, and prediction on a fixed hold-out test phase. The permutations used are defined in Appendix \ref{app:perm}. 

Therefore, for this benchmark, data are gathered from the lower swirl plane (LSP), at 1300 rpm and 40 kPa. \texttt{EngineBench\_LSP\_small} contains 5205 PIV snapshots in total. 

\paragraph{Future development}
The current benchmark intends to serve as a starting point for informing inpainting model selection in engineering applications and investigating model success and failure modes. However, the EngineBench database could be leveraged for a variety of other canonical challenges in flow diagnostics, such as super-resolution or 3D spatial reconstruction. In addition, the development of more general pre-trained models that can be deployed off-the-shelf for a range of industrial flows is of significant interest. The present benchmark will be expanded upon with these goals in mind in order to address a number of challenges. 
\begin{itemize}
    \item Inclusion of additional PIV planes to test predictive capabilities for a range of fields of view.
    \item Inclusion of data from different experimental rigs. 
    \item Test inpainting performance on a range of gap types.
    \item Benchmarking a wider range of standard problems such as super-resolution and 3D reconstruction from 2D data.
\end{itemize}

\begin{figure}[t]
    \centering
    \includegraphics[width=0.3\columnwidth]{fig/snapp.png}\put(-122,90){\large (a)}
    \hspace{3mm}
    \includegraphics[width=0.3\columnwidth]{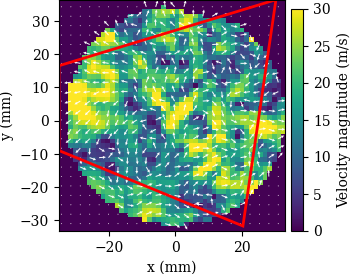}\put(-122,90){\large (b)}
    \hspace{3mm}
    \includegraphics[width=0.3\columnwidth]{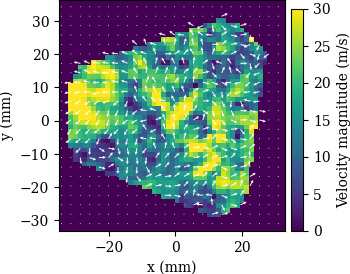}\put(-122,90){\large (c)}
    \caption{Example random edge gap creation. From left to right, (a): original image; (b): image with a random edge gap polygon superimposed in red; (c): edge gaps added to regions outside of the random polygon.}
    \label{fig:traingaps}
\end{figure}

\section{Benchmark setup}
\label{sec:bench}

\begin{wrapfigure}{R}{0.45\textwidth}
\begin{center}
\includegraphics[width=0.45\columnwidth]{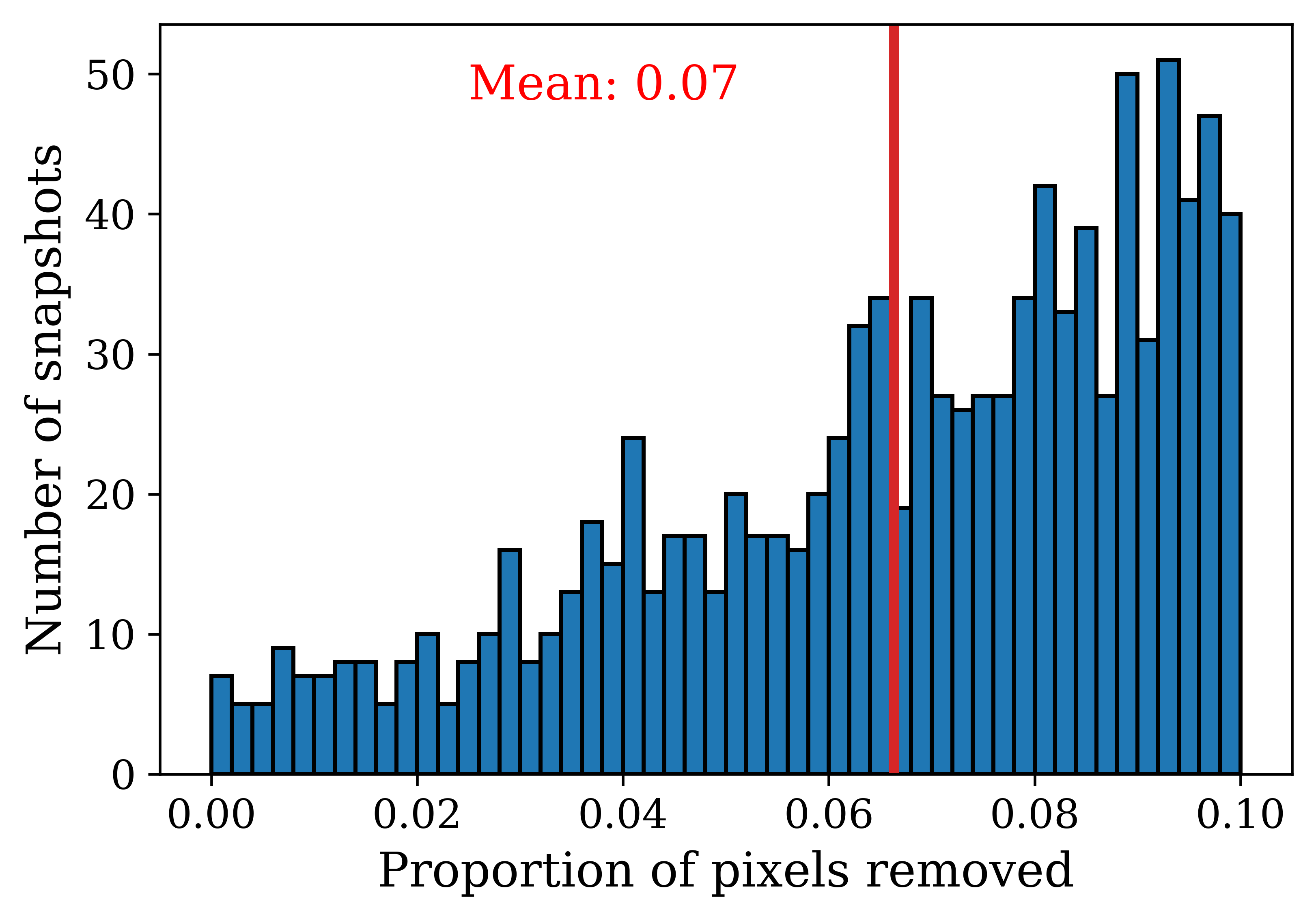}
\end{center}
    \caption{Histogram showing the proportion of pixels removed by the random edge masks in one pass through the training set. 7\% of the total pixels in the field of view were removed on average.}
    \label{fig:hist}
\end{wrapfigure}

\paragraph{Data}
Of the five phase angles in \texttt{EngineBench\_LSP\_small}, the data at one phase angle, 180 crank angle degrees (CAD), are held out for the test case. The flow fields at 180 CAD are notoriously challenging to predict, as the piston is on the point of switching its direction of travel, causing the flow patterns to be highly variable~\cite{ko2021methodology}. Holding out this phase angle is therefore deemed a sufficient test for generalizability within a single test point. Different permutations of the other four phase angles are then used to construct the training and validation sets, with three phases for training and one for validation, as described in Appendix~\ref{app:perm}. The resultant number of images in the train, validation, and test sets are 3123, 1041, 1041. Data at each phase angle are standardized separately by subtracting the phase mean and dividing by the phase standard deviation. Horizontal and vertical velocity components are treated as separate colour channels, and the original spatial dimensions for each image are 50 $\times$ 49 pixels. Zero padding is added around the edges of the images to 128 $\times$ 128 for compatibility with standard ML models.

\paragraph{Gap creation}
During training, edge gaps of random sizes and orientations are added to the training data. The gappy data form the features, and the original full flow fields are the labels. The random edge gaps are constructed by taking four random points along the input image borders, drawing a polygon between the points, and masking out pixels that lie outside of the polygon. There can be a maximum of two points on any one edge. This approach ensures that edge gaps are created with random sizes and orientations, to prevent the models from overfitting to specific gap shapes and locations. This is of increased importance as this benchmark is intended to be used for multi-field of view and multi-geometry predictions in the future, as mentioned in Section \ref{sec:data}.
A regular PIV snapshot is shown alongside two snapshots with random edge gaps added in Figure \ref{fig:traingaps}. A maximum percentage of the pixels are allowed to be removed by the mask; the mask is discarded if it removes more pixels than this, and a replacement mask is generated. This upper threshold for the gap sizes is needed to constrain the training process, prevent the inpainting task from becoming overly challenging, and reflect more realistic physical scenarios. Two gap sizes are chosen for this benchmark: 10\% and 25\%. 
A histogram showing the proportion of pixels removed for each snapshot in one pass of the training set for 10\% gaps is shown in Figure \ref{fig:hist}. For the test data, two masks of fixed shape are constructed that each remove a the required threshold of the pixels at the edges of the field of view. A vertical mask is applied to the first half of the test set, and a horizontal mask to the latter half. 

\paragraph{Models}
The performance of five different modeling setups is benchmarked in this study. Firstly, adaptive median filter GPOD (GPOD-MF) is chosen as a best-in-class non-parametric approach~\cite{saini2016development}. Secondly, the UNet model \cite{ronneberger2015u} is chosen due to its popularity in turbulent flows. Two loss functions are tested with the UNet; a mean square error (MSE) loss, and a huber loss function in order to test the effect of outliers in the PIV data. Further details of the loss functions are given in Appendix \ref{app:loss}. Fourthly, the UNet transformer (UNETR) model \cite{hatamizadeh2022unetr} with a MSE loss is chosen due to the performance enhancements that have been reported due to the transformer module, with the project MONAI implementation \cite{cardoso2022monai}. Finally, an adapted version of the context encoder generative adversarial neural network (CE-GAN) \cite{pathak2016context} with MSE and adversarial losses is implemented due to its high performance in standard inpainting tasks~\cite{elharrouss2020image,liu2021pd} and recent usage in turbulent flows \cite{li2023multi,buzzicotti2021reconstruction}. As the original context encoder was designed for inpainting gaps of a fixed size and location, the network architecture is modified in a similar fashion to Li \textit{et al.}~\cite{li2023multi}. For the generator, an additional \texttt{upconv} layer is included at the output to return a prediction of the same spatial dimensions as the input. To correspond with the generator modifications, an extra \texttt{conv} layer is added at the beginning of the discriminator to handle inputs of the same size as the original data. A dropout layer with $p=50\%$ is also added at the output of the discriminator, following Li \textit{et al.} \cite{li2023multi}.

\paragraph{Inpainting setup} During training, the losses between the model predictions and the labels are calculated across the entire image, not just inside the gap. This approach provides a number of benefits: to simplify the random gaps training process, retain the context of the broader turbulent flow and field of view, and to provide practitioners with a visual representation of how the network relates the prediction inside the gap to the rest of the field, avoiding edge effects in the output. Performance metrics on the test set predictions are then given for the central regions as well as the gappy edge regions. Training hyperparameters are chosen to reflect other turbulent flow ML studies~\cite{chung2024turbulence,li2023multi}. Training in all cases was run over 300 epochs. For the UNet and UNETR models, the learning rate was 1e$-$3 and multiplied by a factor of 0.5 every 50 epochs via a step scheduler. For the CE-GAN, the learning rate was 1e$-$4 and multiplied by 0.75 every 50 epochs. All architectural hyperparameters are maintained from their original studies. The models were trained on an NVIDIA GeForce GTX Titan X GPU. The training process for each model was run three times with different permutations of training and validation phase angles to provide the error bars and standard deviations.  

\paragraph{Metrics} 
The relative L2 error is used to quantify pixelwise accuracy, and is calculated as follows:
\begin{equation}
\textrm{L2} =\left\|\boldsymbol{u}_{\mathrm{true}}-\boldsymbol{u}_{\mathrm{pred}}\right\|_2 /\left\|\boldsymbol{u}_{\mathrm{true}}\right\|_2
\end{equation}

where $\boldsymbol{u}_{\mathrm{true}}$ and $\boldsymbol{u}_{\mathrm{pred}}$ are the true and model-predicted velocity vectors respectively. In addition, two vector-based metrics are used to quantify the similarity of the overall flow structures. For directional similarity, the relevance index (RI) \cite{liu2011development} is calculated as:
\begin{equation}
    \textrm{RI}=\frac{\langle \boldsymbol{u}_{\mathrm{true}},\boldsymbol{u}_{\mathrm{pred}}\rangle}{\left\|\boldsymbol{u}_{\mathrm{true}}\right\|_2  \left\|\boldsymbol{u}_{\mathrm{pred}}\right\|_2}
    \label{eqn:RI}
\end{equation}

where $\langle \cdot, \cdot \rangle$ represents the inner product. The RI varies between $1$ for perfectly aligned vectors, and $-1$ for perfectly opposite vectors. The similarity of the vector magnitudes is given by the magnitude index (MI) \cite{hu2015large}:
\begin{equation}
    \textrm{MI} = 1-\frac{\left\|\boldsymbol{u}_{\mathrm{true}}-\boldsymbol{u}_{\mathrm{pred}}\right\|_2}{\left\|\boldsymbol{u}_{\mathrm{true}}\right\|_2+\left\|\boldsymbol{u}_{\mathrm{pred}}\right\|_2}
\end{equation}
with the MI varying between 1 for vectors of identical magnitude and 0 for totally disparate vector magnitudes. Finally, in order to capture the multi-scale turbulent flow features, the energy spectrum for each image image is calculated using the Fourier transform:
\begin{equation}
    E(\boldsymbol{k}) = \frac{1}{2}\widetilde{\left(\hat{\boldsymbol{u}}(\boldsymbol{k}) \hat{\boldsymbol{u}}^*(\boldsymbol{k})\right)}
\end{equation}
where $\hat{\boldsymbol{u}}(\boldsymbol{k})$ is the Fourier-transformed velocity vector, $\hat{\boldsymbol{u}}^*(\boldsymbol{k})$ is its complex conjugate, $\boldsymbol{k}$ is the spatial frequency wavenumber vector, and $\widetilde{(\cdot)}$ represents the radial average over the vertical and horizontal frequencies \cite{li2023multi}. The Kullback–Leibler (KL) divergence is then used to quantify the similarity between energy spectra:
\begin{equation}
    \textrm{KL}(E_{\mathrm{true}} \| E_{\mathrm{pred}})=\sum_k E_{\mathrm{true}}(k) \log \left(\frac{E_{\mathrm{true}}(k)}{E_{\mathrm{pred}}(k)}\right)
\end{equation}
ranging from 0 for identical distributions to infinity for a complete divergence.

\section{Results and discussion}
\begin{table}[t]
  \caption{Results for the 180 CAD test case. The mean and standard deviations are reported from the three permutations of training data defined in Appendix \ref{app:perm}. \textbf{Bold} typeface represents the best mean in each category separated by the horizontal lines. GPOD-MF metrics are not given in the central regions as adaptive GPOD methods only update values in the gap locations.}
  \label{tab:results}
  \centering
  \footnotesize
  \begin{tabular}{lccccc}
    \toprule
         & Gap size & RI     & MI  & L2 & KL \\
    \midrule
    Central regions: & & & & & \\
    \; UNet, MSE & 10\% & \textbf{1.000 $\pm$ 0.000} & 0.984 $\pm$ 0.000 & 0.033 $\pm$ 0.001 & \textbf{0.000 $\pm$ 0.000} \\
    \; UNet, huber & 10\% & 0.999 $\pm$ 0.000 & 0.983 $\pm$ 0.000  &  0.035 $\pm$ 0.001  & \textbf{0.000 $\pm$ 0.000} \\
    \; UNETR & 10\% & \textbf{1.000 $\pm$ 0.000} & \textbf{0.985 $\pm$ 0.001} & \textbf{0.030 $\pm$ 0.001}  & \textbf{0.000 $\pm$ 0.000} \\
    \; CE-GAN & 10\% & 0.884 $\pm$ 0.003 & 0.745 $\pm$ 0.010 & 0.470 $\pm$ 0.008  & 0.019 $\pm$ 0.004 \\
    \midrule
    Edge gaps: &&&&&\\
    \; GPOD-MF & 10\% & 0.797 $\pm$ 0.001 & 0.666 $\pm$ 0.001 & 0.610 $\pm$ 0.002 & 0.105 $\pm$ 0.001 \\
    \; UNet, MSE & 10\% & 0.890 $\pm$ 0.004 & 0.759 $\pm$ 0.007 & 0.456 $\pm$ 0.009 & \textbf{0.013 $\pm$ 0.002} \\
    \; UNet, huber & 10\% & \textbf{0.892 $\pm$ 0.003} & \textbf{0.760 $\pm$ 0.003}  &  \textbf{0.452 $\pm$ 0.005}  & \textbf{0.013 $\pm$ 0.001} \\
    \; UNETR & 10\% & 0.884 $\pm$ 0.002 & 0.755 $\pm$ 0.002 & 0.467 $\pm$ 0.005  & 0.014 $\pm$ 0.001 \\
    \; CE-GAN & 10\% & 0.784 $\pm$ 0.008 & 0.661 $\pm$ 0.010 & 0.622 $\pm$ 0.007  & 0.032 $\pm$ 0.008 \\
    \midrule 
    Central regions: & & & & & \\
    \; UNet, MSE & 25\% & \textbf{0.999 $\pm$ 0.000} & 0.978 $\pm$ 0.001 & 0.044 $\pm$ 0.002 & \textbf{0.000 $\pm$ 0.000} \\
    \; UNet, huber & 25\% & \textbf{0.999 $\pm$ 0.000} & 0.978 $\pm$ 0.001  &  0.045 $\pm$ 0.002  & \textbf{0.000 $\pm$ 0.000} \\
    \; UNETR & 25\% & \textbf{0.999 $\pm$ 0.000} & \textbf{0.983 $\pm$ 0.001} & \textbf{0.034 $\pm$ 0.003}  & \textbf{0.000 $\pm$ 0.000} \\
    \; CE-GAN & 25\% & 0.885 $\pm$ 0.006 & 0.739 $\pm$ 0.006 & 0.470 $\pm$ 0.011  & 0.020 $\pm$ 0.001 \\
    \midrule
    Edge gaps: &&&&&\\
    \; GPOD-MF & 25\% & 0.762 $\pm$ 0.008 & 0.629 $\pm$ 0.007 & 0.654 $\pm$ 0.009 & 0.144 $\pm$ 0.011 \\
    \; UNet, MSE & 25\% & 0.817 $\pm$ 0.005 & \textbf{0.691 $\pm$ 0.006} & 0.571 $\pm$ 0.006 & 0.029 $\pm$ 0.001 \\
    \; UNet, huber & 25\% & \textbf{0.822 $\pm$ 0.001} & \textbf{0.691 $\pm$ 0.004}  &  \textbf{0.565 $\pm$ 0.002}  & 0.028 $\pm$ 0.001 \\
    \; UNETR & 25\% & 0.800 $\pm$ 0.005 & 0.680 $\pm$ 0.004 & 0.598 $\pm$ 0.004  & \textbf{0.027 $\pm$ 0.001} \\
    \; CE-GAN & 25\% & 0.735 $\pm$ 0.005 & 0.620 $\pm$ 0.004 & 0.677 $\pm$ 0.005  & 0.055 $\pm$ 0.004 \\
    \bottomrule
  \end{tabular}
\end{table}

The results for the benchmark performance metrics are given in Table \ref{tab:results}, with the best result for each metric presented in bold. The UNet and UNETR models exhibit similar performances across all metrics, with the UNet models slightly outperforming UNETR for predictions inside the edge gaps. As shown in Appendix \ref{app:modelsizes}, the number of parameters in the UNet architecture is just 12\% of that of the UNETR model, so the UNet exhibits a better accuracy-complexity trade-off. This indicates that detailed local features and textures may be more predictive of the target outputs than global context in this situation, which runs counter to where UNETR models typically see performance gains \cite{kang2023new,jiang2023transcfd,xu2023estimation}. The huber loss function with delta = 1 slightly outperforms the MSE loss function on average, but the performance differences are small and within the uncertainty ranges given by the standard deviations. 

\begin{wrapfigure}{R}{0.45\textwidth}
\begin{center}
\includegraphics[width=0.45\columnwidth]{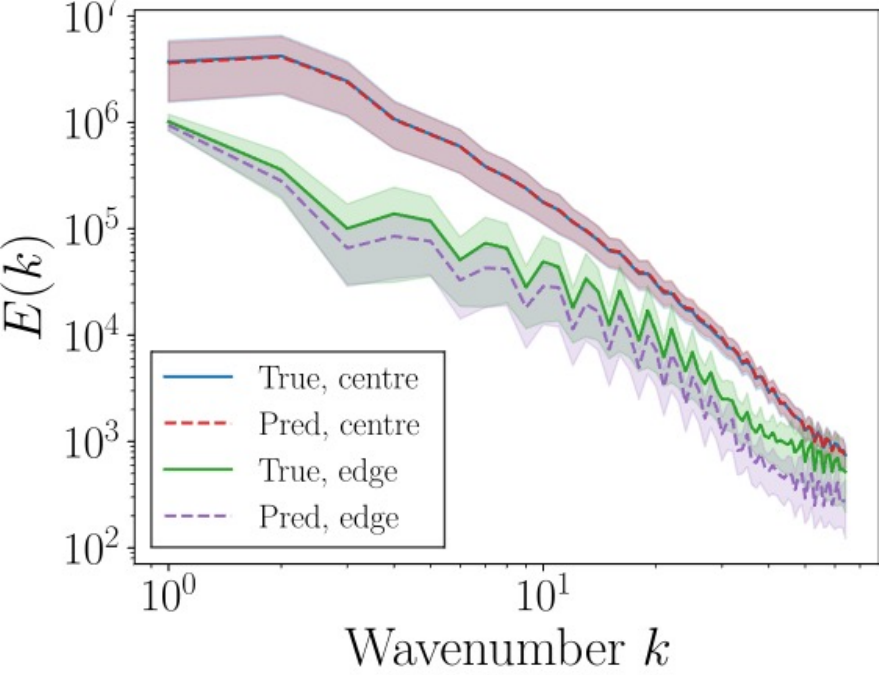}
\end{center}
    \caption{Energy spectra comparing the ground truth test set images to the UNet, MSE predictions at a 10\% test gap size. Ensemble mean spectra are given by solid or dashed lines, with the shaded areas representing one standard deviation from the means.}
    \label{fig:spectra}
\end{wrapfigure}

The accuracy of all UNet-based models is very high in the image centres, with KL divergences that round to zero at a three decimal place tolerance, showing that the original flow structures across all scales are being well-preserved. Ensemble averaged energy spectra for the UNet, MSE model predictions at 10\% gaps are shown in Figure~\ref{fig:spectra}, and there is a near line-on-line match between the true and predicted spectra in the image centres. Concerning predictions in the edge gaps, the relative L2 errors of both UNet and UNETR models are higher at between 45-47\%. This is within the range of values reported by Li \textit{et al.} \cite{li2023multi}, but about twice as high as other results reported by Morimoto \textit{et al.} \cite{morimoto2021experimental} for the reconstruction of a turbulent flow in a fixed gap shape. The latter discrepancy reflects the additional complexity of inpainting large edge gaps of random shape and orientation. 

For the RI and MI, values of between 0.9$-$0.95 are commonly taken to represent self-similarity between vector fields \cite{rulli2021critical}. The average RIs for the UNet and UNETR predictions at 10\% gap sizes approach this criterion in the edge gaps, and meet it in the central regions. The MI values are systematically lower, which is consistent with other reports that the MI is a stricter metric to satisfy as it follows a linear relationship rather than the sinusoidal RI \cite{rulli2021critical,baker2023dynamic,barbato2023cold}. The energy spectra are challenging to compute in the gappy regions in isolation, as sharp edges and discontinuities are prevalent, contributing to the Gibbs phenomena observed in the edge spectra in Figure \ref{fig:spectra}. However, overall trends can still be seen, and the UNet edge prediction follows a downward trend that is similar to the ground truth. 

\begin{figure}[b]
    \centering
    \includegraphics[width=0.3\columnwidth]{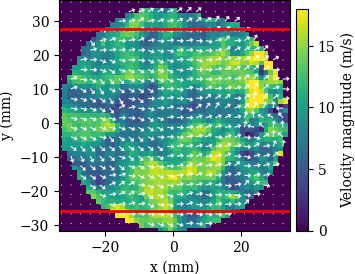}
    \put(-122,82){\large (a)}
    \includegraphics[width=0.3\columnwidth]{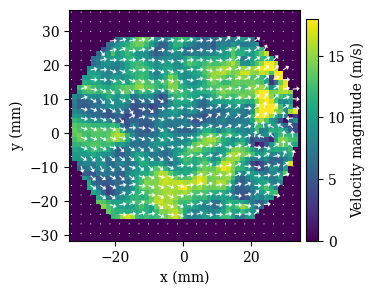}
    \put(-122,82){\large (b)}
    \includegraphics[width=0.3\columnwidth]{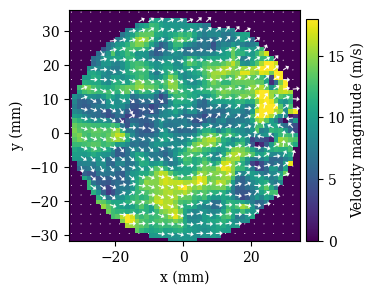}
    \put(-122,82){\large (c)}\\
    \includegraphics[width=0.3\columnwidth]{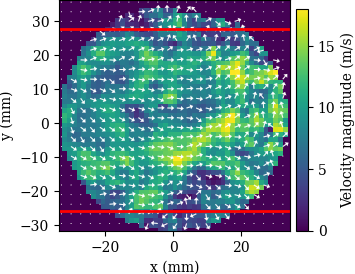}
    \put(-122,82){\large (d)}
    \includegraphics[width=0.3\columnwidth]{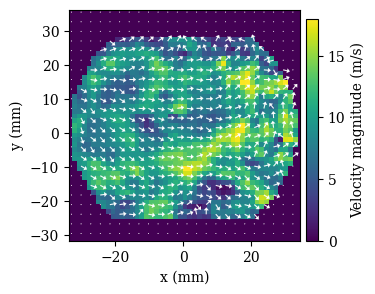}
    \put(-122,82){\large (e)}
    \includegraphics[width=0.3\columnwidth]{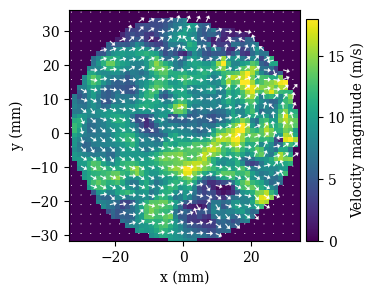}
    \put(-122,82){\large (f)}
    \caption{Sample flow fields from the 10\% gaps test set. Top row: best UNet, MSE prediction (L2 = 0.225); bottom row: worst UNet, MSE prediction (L2 = 1.026). Part (a) and (d): original snapshot with the test mask shown as horizontal red lines; (b) and (e): gappy input; (c) and (f): prediction.}
    \label{fig:results}
\end{figure}

Example flow field predictions from the UNet, MSE model at 10\% gaps are shown in Figure \ref{fig:results}, with predictions from the other models and gap size given in Appendix \ref{app:preds}.
In the top row of Figure \ref{fig:results}, the regions masked out by the horizontal mask are relatively uniform and easy to predict with no large variations in velocity magnitude. On the other hand, for the bottom row, the flow direction switches to point downwards just inside the mask region. There are no obvious indicators for this motion in the rest of the image, and the UNet predicts a simpler left-to-right motion. The lack of spatially local information due to the edge gaps highlights the challenge presented by inpainting in the edge gaps; it is likely that more knowledge of the out-of-plane motion would be needed in order to predict such complex behaviour.

The CE-GAN demonstrates relatively poorer performance across the board. Li \textit{et al.}~\cite{li2023multi} also reported relatively low pixel-wise accuracies for the CE-GAN in an inpainting task on PIV data, but better performance than GPOD in terms of predicting multi-scale properties. These findings are supported in the present study; however, Table \ref{tab:results} shows that the CE-GAN results are worse than UNet-based models, especially in the central image regions. The low central region accuracies indicate that the CE-GAN is not retaining as much of the image centres as the UNet-based models are. Indeed, while the CE-GAN generator has an AlexNet-like structure \cite{krizhevsky2012imagenet}, UNet-based models have skip connections designed to preserve contextual information. This allows the latter to simultaneously preserve information in the image centre to a very high degree of accuracy, and yield gap predictions that seamlessly integrate with the rest of the image. This also correlates with better edge gap predictions for the UNet-based models in this case. It should be noted that this particular application pushes the CE-GAN beyond its initial design intention of solely predicting inside the gap region, rather than also reproducing the entire image. 

Finally, the GPOD-MF method yields the lowest performance, as GPOD-MF is sensitive to the limited amount of spatial information available near the gap regions. Difficulties arise because the algorithm initialises the gaps with ensemble mean vectors calculated from the training set, then iterates on these guesses using the dominant flow features from POD-based reconstructions. However, for highly variational flows, the mean can be a poor approximation of the full dataset \cite{baker2023dynamic}. GPOD-MF can overcome this by utilising the local spatial information to update the guesses, but this is not so effective for large block gaps, and errors can be compounded instead. Figure \ref{fig:gpod_conv} in Appendix \ref{app:gpod} shows that the algorithm converges at a relatively small number of modes, representing the dominant flow structures. While these dominant structures do not fare as badly on the global vector-based metrics, they are overly smoothed and differ vastly in terms smaller-scale flow structures, as shown by the large KL divergences between the GPOD-MF predictions and the true vectors in the edge gaps.

\section{Conclusion}
\label{sec:concl}
This work introduces the EngineBench database to address the limited availability of practical benchmarks on experimental data.
EngineBench is used to benchmark five inpainting methods on edge gaps, which are a novel method of artificial gap creation relevant to internal flow PIV data. The lack of spatially local information due to the edge gaps proves to be an informative challenge for the models, with fewer in-plane indicators of complex flow behaviour. We find that the inclusion of a transformer did not improve UNet predictions in this case, possibly due to small dataset sizes and high snapshot-to-snapshot variations causing local features to be more predictive than global context~\cite{baker2023dynamic}, although further investigation is required. UNet-based models are shown to be more effective than the CE-GAN in this benchmark, due to better retention of local features via the skip connections. UNet-based model predictions approach self-similarity in the 10\% edge gaps according to the vector-based metrics, although pixel-wise L2 errors remain higher than standard computer vision benchmarks. With the increase in gap size, the accuracy of the predictions decrease, although the relative superiority of UNet-based models remains consistent, indicating a degree of robustness in these findings that can be further tested in future work.

\paragraph{Limitations}
This benchmarking study is restricted to a single geometry, PIV plane, and test point, and the modelling results may therefore change with the inclusion of more data. We plan to address this in the very near future by adding more diverse data sources to EngineBench. Furthermore, as the PIV data represent 2D slices of a 3D system, there is limited information about out-of-plane motion which may cause seemingly unexpected behaviour in the PIV data \cite{wieneke2015piv}. 
In addition, edge gaps are the only type of gap considered in this study. 
To date, little work has been
done on quantifying and classifying the types of large structural gaps commonly seen in PIV data from internal flows, and it is possible that model behaviour would vary for different gap types. Finally, although this work focussed on the most well-studied models in turbulent flow inpainting~\cite{li2023multi}, it is possible that performances could be improved with more advanced GAN architectures, or inpainting-adapted neural operators or GNNs, which will also be investigated in future work. 

\clearpage
\begin{ack}

The authors would like to express warm thanks to Kay Song and Christopher Nicholls of the University of Oxford and Wai Tong Chung of Stanford University for their support and advice on this work. This research was funded in whole or in part by the Engineering and Physical Sciences Research Council Prosperity Partnership, Grant No. EP/T005327/1. For the purpose of Open Access, the author has applied a CC BY public copyright license to any Author Accepted Manuscript (AAM) version arising from this submission. The Prosperity Partnership is a collaboration among JLR, Siemens Digital Industries Software, the University of Bath, and the University of Oxford. The authors would also like to thank the Department of Engineering Science technicians and maintenance teams for facilities support. XiaoHang Fang gratefully acknowledges the financial support from the University of Calgary Transdisciplinary Scholarship Connector Grants and the John Fell Oxford University Press Research Fund.
\end{ack}

\small

\bibliographystyle{IEEEtranN}
\bibliography{bib}

\clearpage 
\appendix

\section{Gappy POD method}
\label{app:gpod}
\paragraph{GPOD algorithm}
The median filter adaptive GPOD algorithm used in this paper was developed by Saini \textit{et al.} \cite{saini2016development}. Like all POD methods, GPOD uses the singular value decomposition (SVD) to find a set of basis functions that allow dominant features in the data to be ordered and extracted. POD-based reconstructions of the data can then be produced via linear combinations of these basis functions. For more details on the SVD and POD, see Refs. \cite{taira2017modal,brunton2021data}. Following the notation used by Saini \textit{et al.} \cite{saini2016development}, we consider a field $\xi(\mathbf{x}, t)$ defined by spatial measurement vectors $\mathbf{x}$ taken at discrete points in time $t$. The field's POD-based reconstruction $\breve{\xi}$ can be written as a linear combination of temporal coefficients $a_i$ and spatial modes $\phi_i$:
$$
\left.\breve{\xi}\left(\mathbf{x}, t_k\right)\right|_P=\sum_{i=1}^P a_i^k \phi_i(\mathbf{x}) ; \quad k \in[1, N]
$$

where $P$ is the number of modes and $N$ is the number of snapshots. The following algorithm is then applied, consisting of a nested loop of iterations until convergence, with sub iterations at a fixed number of POD modes indexed by $j$ and main iterations where the number of POD modes is incremented given by $n$. The algorithm is iterative in nature because the optimal number of POD modes to be included in the reconstructions is not known in advance, and the inclusion of too few or too many POD modes is akin to under- or over-fitting the data. Like the original studies, we start with two POD modes in the reconstruction and increment the number of modes one at a time until main loop convergence \cite{saini2016development,venturi2004gappy}.
\begin{enumerate}
    \item To initialise the algorithm, all gaps in the input field are replaced by the ensemble mean at the corresponding spatial location, to give the filled field $\tilde{\xi}_{0,0}$: \\
    $$
\tilde{\xi}_{0,0}\left(\mathbf{x}, t_k\right)= \begin{cases}\xi\left(\mathbf{x}, t_k\right) & \left.\mathbf{x} \in \mathbf{x}_{\mathrm{d}}\right|_{t_k} \\ \Bar{\xi}(\mathbf{x}) & \left.\mathbf{x} \in \mathbf{x}_{\mathrm{g}}\right|_{t_k}\end{cases}
$$
where $\mathbf{x}_{\mathrm{d}}$ and $\mathbf{x}_{\mathrm{g}}$ represent the locations of data points and gaps respectively, and $\Bar{\xi}(\mathbf{x})$  is the ensemble mean. Note that the algorithm cannot initialise if a gap exists in all snapshots.  

\item At a given main iteration $n$, POD is performed on $\tilde{\xi}_{n,j}$ using $P_n$ modes, resulting in a POD-reconstructed approximation $\breve{\breve{\xi}}$:
$$
\breve{\breve{\xi}}_{n,j}\left(\mathbf{x}, t_k\right)=\sum_{i=1}^{P_n} a_i^k \phi_i(\mathbf{x})
$$

\item The gap locations in the filled field are then updated with values from the POD approximation.
$$
\tilde{\xi}_{n, j+1}\left(\mathbf{x}, t_k\right)= \begin{cases}\xi\left(\mathbf{x}, t_k\right) & \left.\mathbf{x} \in \mathbf{x}_{\mathrm{d}}\right|_{t_k} \\ \breve{\breve{\xi}}_{n, j}\left(\mathbf{x}, t_k\right) & \left.\mathbf{x} \in \mathbf{x}_{\mathrm{g}}\right|_{t_k}\end{cases}
$$

\item Steps (2$-$3) are repeated until the POD eigenvalues computed in step 2 converge to within a user-defined tolerance. The final approximation at main iteration $n$ is retained as $\hat{\xi}_{n}$.

\item A median filter (MF) is then applied as an outlier detection technique in order to adaptively retain promising guesses and revert poor guesses to their previous values. The MF selection is implemented by calculating the residual $R$ of a centre pixel $\mathbf{x}_{\mathrm{cp}}$ in a neighborhood of adjacent pixels $\mathbf{x}_{ad}$.
$$
\left.R(\xi)\right|_{\mathbf{x}_{\mathrm{cp}}, t_k}=abs\left(\xi\left(\mathbf{x}_{\mathrm{cp}}, t_k\right)-\operatorname{median}\left(\xi\left(\mathbf{x}_{ad}, t_k\right)\right)\right)
$$

The updated guesses at each pixel are retained if the residual at that pixel is reduced relative to the previous main iteration, and reverted otherwise:
$$
\hat{\xi}_n\left(\mathbf{x}, t_k\right)= \begin{cases}\hat{\xi}_{n-1}\left(\mathbf{x}_{\mathrm{cp}}, t_k\right), & \text { if }\left.R_{n-1}\right|_{\mathbf{x}_{\mathrm{cp}}, t_k} \leq\left. R_n\right|_{\mathbf{x}_{\mathrm{cp}}, t_k} \\ \hat{\xi}_n\left(\mathbf{x}_{\mathrm{cp}}, t_k\right), & \text { if }\left.R_{n-1}\right|_{\mathbf{x}_{\mathrm{cp}}, t_k}>\left.R_n\right|_{\mathbf{x}_{\mathrm{cp}}, t_k}\end{cases}
$$
and the next main iteration begins with $\tilde{\xi}_{n+1,0}$ = $\hat{\xi}_n$.

\item The algorithm increments the main iterations $n$ until a main convergence criterion is satisfied, as discussed below.

\end{enumerate}

\paragraph{Main convergence criterion}
In practical scenarios, the true values of the data in the edge gaps would not be known, so the accuracy of the GPOD reconstructions cannot be assessed directly. A convergence criterion is therefore needed in order to prevent over-fitting and terminate the algorithm at a number of modes that is close to the true optimal value.
A commonly-used criterion was introduced by Gunes \textit{et al.} \cite{gunes2006gappy} known as convergence checking (CC) gaps. With this method, additional gaps are added to the input data in locations where the true values are known. The reconstruction error of the GPOD reconstructions inside these CC gaps can then be calculated and tracked as a proxy for the true reconstruction errors inside the real gaps. The GPOD algorithm is terminated when the CC reconstruction errors are deemed to have hit a minimum.

\begin{figure}[h]
    \centering 
    \includegraphics[width=0.45\columnwidth]{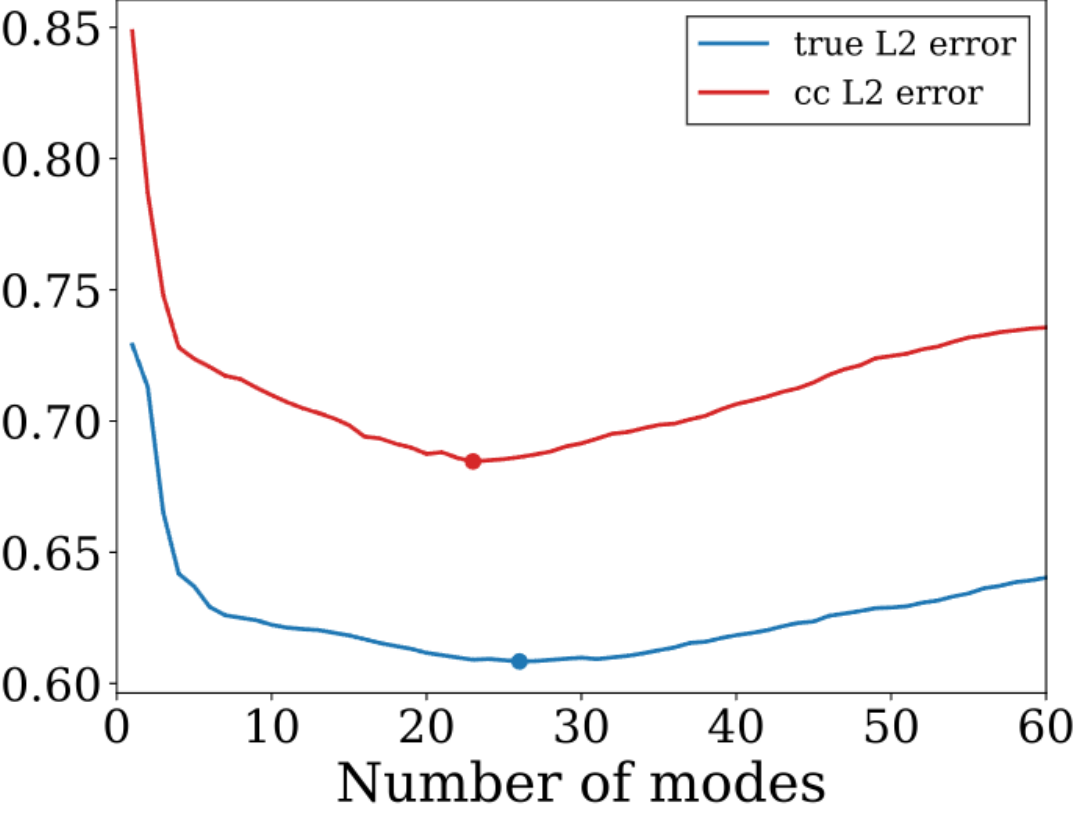}\hspace{5mm}
    \includegraphics[width=0.45\columnwidth]{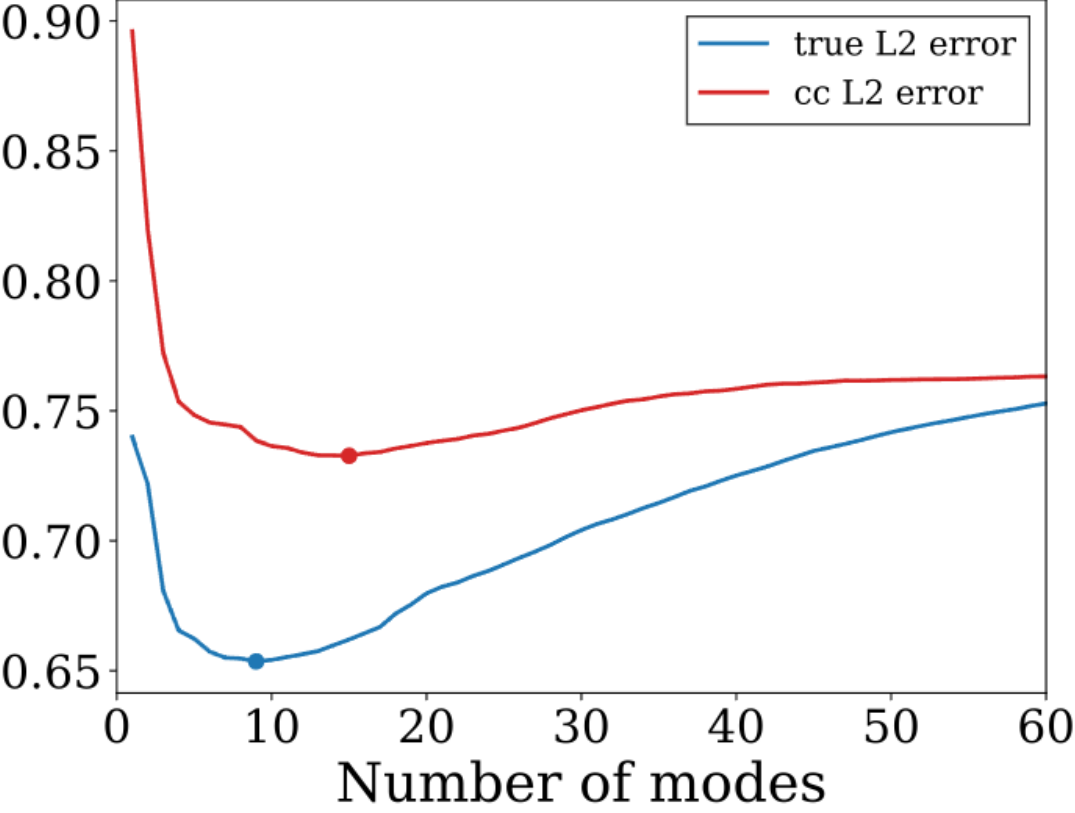}
    \caption{GPOD convergence curves for 10\% and 25\% gaps (left and right, respectively) at permutation A. The minimum errors for both the true L2 error in the edge gaps and the L2 error in the convergence checking (cc) gaps are marked as filled circles.}
    \label{fig:gpod_conv}
\end{figure}

\paragraph{Implementation}
GPOD is performed on a single large data matrix, rather than separate training and testing matrices. Therefore, in the present study, the test data matrix at 180 CAD is stacked alongside three training data matrices according to the permutations defined in Appendix \ref{app:perm}. The test edge gaps are added to the test data matrix, and the gap locations are initialised using the ensemble mean from the training data matrices which do not contain any gaps. CC gaps are also implemented into the test data in a similar format to the test edge gaps; ie, the test edge gaps are extended to incorporate another 10\% of the pixels in the image where the true values are known. The GPOD-reconstructed flow fields at the minimum CC L2 error are retained for analysis in this study. Plots showing the GPOD convergence curves for 10\% and 25\% gaps in permutation A are shown in Figure \ref{fig:gpod_conv}. In both cases, the relative L2 error calculated in the CC gaps gives an optimal number of modes that is relatively close to the true optimum given by the L2 error in the true gaps. 

\section{Engine operation details}
\label{app:exp}
The TCC-III is a single-cylinder spark-ignition transparent research engine. As the piston moves, the engine operation follows a standard four-stroke cycle, with intake, compression, expansion, and exhaust strokes. 
A schematic of the engine geometry is shown in Figure \ref{fig:schematic}; the cylinder has a diameter of 92.0 mm and an effective height of between 9.5 mm and 86.0 mm depending on the position of the piston. Further geometry and valve timing details are given in Table \ref{tab:valves}. Data currently used in EngineBench were taken from motored (i.e. operated by a dynamometer with no fuel injected) runs of the engine at the operating points given in Table \ref{tab:op}. 

\begin{table}[h]
\caption{TCC-III engine geometry and valve details. Valve opening and closing events are given in crank angle degrees after top dead-centre exhaust (CAD aTDCexh).}
\label{tab:valves}
\centering
\begin{tabular}{ll}
\toprule
\textbf{Parameter}                 & \textbf{Value} \\ \midrule
Bore [cm]                           & 9.20           \\
Stroke [cm]                         & 8.60           \\
Swept volume [cc]                      & 571.7          \\
Compression ratio [$-$] & 10:1 \\
Exhaust valve closing [CAD aTDCexh] & 12.8           \\
Intake valve closing [CAD aTDCexh]  & 240.8          \\
Exhaust valve opening [CAD aTDCexh] & 484.8          \\
Intake valve opening [CAD aTDCexh]  & 712.8          \\ \bottomrule
\end{tabular}
\end{table}

\begin{table}[h]
\caption{TCC-III engine operating points used in EngineBench.}
\label{tab:op}
\centering
\begin{tabular}{ccc}
\hline
\textbf{Operating point} & \textbf{Intake air pressure {[}kPa{]}} & \textbf{Engine speed {[}rpm{]}} \\ \hline
A                        & 95                                     & 800                             \\
B                        & 95                                     & 1300                            \\
C                        & 40                                     & 1300                            \\ \hline
\end{tabular}
\end{table}


Optical access is achieved via a fully quartz cylinder and a flat quartz piston window with a diameter of 70 mm. To conduct PIV, the airflow was seeded with silicone oil droplets with a mean diameter of 1 $\mu$m. The flow along a given PIV plane was then illuminated by a high-repetition-rate dual-cavity frequency-doubled ND:YLF laser, and imaged by a high-speed Phantom v1610 camera. The laser pulse separation varied between 3$-$80 $\mu$s depending on the phase angle and the speed of the flow. Vector fields were generated from the raw images using the LaVision DaVis 8.x software, with final spatial resolutions between the vectors of 1.25$-$1.4 mm. For more details on the experimental apparatus see Ref. \cite{schiffmann2016tcc}.

\section{HDF5 file structure}
\label{app:h5}

\begin{figure}[h]
    \centering    
    \includegraphics[width=0.7\textwidth]{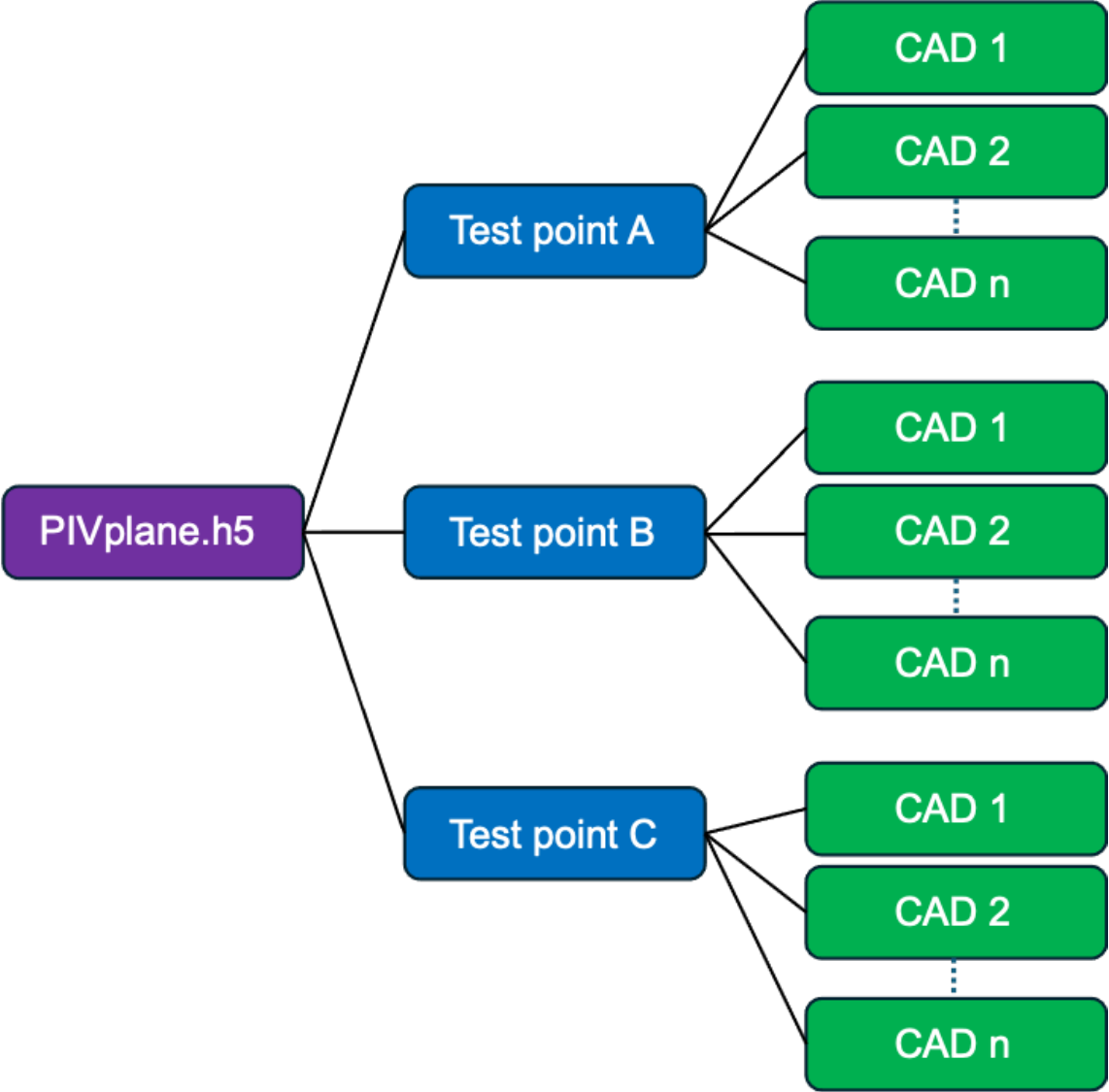}
    \caption{Generalised HDF5 file structure in EngineBench.}
    \label{fig:data_tree}
\end{figure}

The PIV data given in EngineBench are structured as HDF5 files in order to facilitate chunking of the data into different training permutations. A hierarchical structure is implemented, with each HDF5 file containing the data from a single PIV plane, as shown in Figure \ref{fig:data_tree}.

\clearpage
\section{Permutation definition}
\label{app:perm}
The HDF5 file structure is leveraged to enable the training and validation datasets to be comprised of different permutations of individual phase angles. This tests the sensitivity of the model performances to the specific phases chosen for the analysis and provides the error bars for the results. The turbulence dynamics vary significantly between different phase angles during the engine cycle, with the working fluid undergoing rapid compressions and expansions; training models on different phase angles therefore exposes the models to different turbulent dynamics and serves as an initial step in testing model generalisability. For the given number of phase angles considered, only three permutations are considered out of the possible four as the spread across permutations was found to be acceptably low on all metrics.
\begin{table}[h]
  \caption{Definitions of phase angle permutations that comprise the training, validation, and hold-out test sets. The different permutations are denoted as \textbf{A}, \textbf{B} and \textbf{C}, and the corresponding phase angles are given in crank angle degrees (CAD).}
  \label{tab:perm}
  \centering
  \begin{tabular}{lccc}
    \toprule
         & \textbf{A}     & \textbf{B}  & \textbf{C}  \\
    \midrule
    Train  & 90  & 135 &  90 \\
              & 135 & 225 & 225 \\
              & 225 & 270 & 270 \\
    \midrule
    Validation & 270 & 90 & 135 \\
    \midrule
    Test  & 180 & 180 & 180 \\
    \bottomrule
  \end{tabular}
\end{table}

\section{Loss functions and hyperparameters}
\label{app:loss}

All loss functions were implemented using PyTorch 2.3 functions. The mean-squared-error (MSE) loss $l_{mse}$ between two images is given by:
$$
l_{mse} = \frac{1}{N} \sum_{i=1}^{N}\left( x_i - y_i \right)^2$$
for features $x$, labels $y$, and number of pixels $N$. 

The Huber loss is a hybrid loss function that reduces sensitivity to outliers by applying an L1 loss to element-wise errors above a certain threshold (delta) and a quadratic loss otherwise to aid convergence. It is defined per pixel $i$ as:
$$
l_{huber,i}= \begin{cases}0.5\left(x_i-y_i\right)^2, & \text { if }\left|x_i-y_i\right|<\text { delta } \\ \operatorname{delta} *\left(\left|x_i-y_i\right|-0.5 * \text { delta }\right), & \text { otherwise }\end{cases}
$$
then averaged over all pixels in the image pairing. A smaller value of delta increases the influence of the L1 loss; the value of delta was tuned via a grid search of values presented in Table \ref{tab:huber}. Delta = 1 displayed the best performance in the edge gaps and was therefore used in the remainder of the study.

For the CE-GAN, the discriminator was trained using a binary cross entropy (BCE) loss, while the generator utilised a BCE / MSE hybrid. The BCE loss is defined as:
$$
l_{bce}=-y * \log x+\left(1-y\right) * \log \left(1-x\right)
$$
while the combined generator loss is given by:
$$
l_{gen} = \lambda_{adv} * l_{bce} + (1 - \lambda_{adv}) * l_{mse} 
$$
where the adversarial ratio $\lambda_{adv}$ controls the relative importance of the MSE and BCE losses. Following Li \textit{et al.}~\cite{li2023multi}, we test the sensitivity of four different adversarial ratios given in Table \ref{tab:adver}. $\lambda_{adv} = 1e-2$ demonstrated the best performance across the four metrics and was therefore chosen for the remainder of the CE-GAN results in this study.

\begin{table}[h]
  \caption{Huber loss delta tuning results for the 180 CAD test case with 10\% edge gaps. One result for each setup using permutation A is reported. \textbf{Bold} typeface represents the best result.}
  \label{tab:huber}
  \centering
  \footnotesize
  \begin{tabular}{lcccc}
    \toprule
         & RI     & MI  & L2 & KL \\
    \midrule
    Central regions: & & & & \\
    \; Delta = 5 & 0.999 & 0.983 & 0.033 & \textbf{0.000} \\
    \; Delta = 1 & \textbf{1.000} & 0.983 & 0.034 & \textbf{0.000} \\
    \; Delta = 0.5 & 0.999 & 0.983 & 0.034   & \textbf{0.000} \\
    \; Delta = 0.1 & \textbf{1.000} & \textbf{0.988} & \textbf{0.024} & \textbf{0.000} \\
    \midrule
    Edge gaps: &&&&\\
    \; Delta = 5 & 0.886 & 0.751 & 0.462 & 0.016 \\
    \; Delta = 1 & \textbf{0.896} & \textbf{0.765} & \textbf{0.445} & \textbf{0.013} \\
    \; Delta = 0.5 & 0.895 & 0.760 & 0.449 & \textbf{0.013} \\
    \; Delta = 0.1 & 0.858 & 0.712 & 0.517 & 0.026 \\
    \bottomrule
  \end{tabular}
\end{table}

\begin{table}[h]
  \caption{Adversarial loss lambda tuning results for the 180 CAD test case with 10\% edge gaps. One result for each setup using permutation A is reported. \textbf{Bold} typeface represents the best result.}
  \label{tab:adver}
  \centering
  \footnotesize
  \begin{tabular}{lcccc}
    \toprule
         & RI     & MI  & L2 & KL \\
    \midrule
    Central regions: & & & & \\
    \; Lambda = 1e$-$1 & 0.709 & 0.578 & 0.709 & 0.099 \\
    \; Lambda = 1e$-$2 & \textbf{0.883} & \textbf{0.731} & \textbf{0.478} & \textbf{0.014} \\
    \; Lambda = 1e$-$3 & 0.709 & 0.616 & 0.809 & 0.082 \\
    \; Lambda = 1e$-$4 & 0.773 & 0.645 & 0.807 & 0.024 \\
    \midrule
    Edge gaps: &&&&\\
    \; Lambda = 1e$-$1 & 0.575 & 0.513 & 0.817 & 0.160 \\
    \; Lambda = 1e$-$2 & \textbf{0.789} & \textbf{0.656} & \textbf{0.612} & 0.043 \\
    \; Lambda = 1e$-$3 & 0.695 & 0.604 & 0.855 & \textbf{0.019} \\
    \; Lambda = 1e$-$4 & 0.755 & 0.616 & 0.929 & 0.021 \\
    \bottomrule
  \end{tabular}
\end{table}


\newpage
\section{Model sizes}
\label{app:modelsizes}

Table \ref{tab:times} contains the training times for the ML models using a single NVIDIA GeForce GTX Titan X GPU over 300 epochs. Note that the training times are similar across the different models; this is because the key computational bottleneck is the generation of random masks on the fly. Alternative methods of introducing gaps to the data are currently under investigation, such as selecting masks from a pre-generated set.

\begin{table}[h]
  \caption{ML model training times, and model sizes in millions of parameters.}
  \label{tab:times}
  \centering
  \footnotesize
  \begin{tabular}{lcc}
    \toprule
    Model     & Training time (hrs)     &  \# Parameters (M) \\
    \midrule
    UNet & 19.1 $\pm$ 0.3 & 10.5 \\
    UNETR & 18.7 $\pm$ 0.2  & 87.3 \\
    CE-GAN & 18.5 $\pm$ 0.3 & 74.0 \\
    \midrule 
    Total (all model variations) & $\approx$ 25 days & \\
    \bottomrule
  \end{tabular}
\end{table}


\clearpage
\section{Model predictions}
\label{app:preds}
\begin{figure}[htbp]
    \centering
    \begin{subfigure}[b]{0.95\textwidth}
    \centering
        \begin{subfigure}[b]{0.27\textwidth}
            \includegraphics[width=\textwidth]{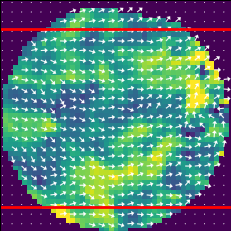}
            \caption*{Original}
        \end{subfigure}
        \begin{subfigure}[b]{0.27\textwidth}
            \includegraphics[width=\textwidth]{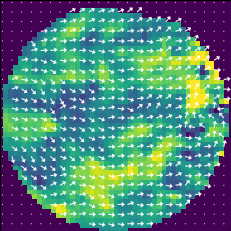}
            \caption*{UNet, MSE}
        \end{subfigure}
        \begin{subfigure}[b]{0.27\textwidth}
            \includegraphics[width=\textwidth]{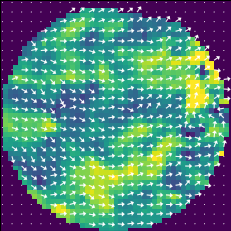}
            \caption*{UNet, huber}
        \end{subfigure}\\
        \begin{subfigure}[b]{0.27\textwidth}
            \includegraphics[width=\textwidth]{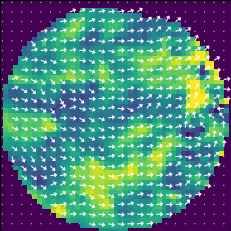}
            \caption*{UNETR}
        \end{subfigure}
        \begin{subfigure}[b]{0.27\textwidth}
            \includegraphics[width=\textwidth]{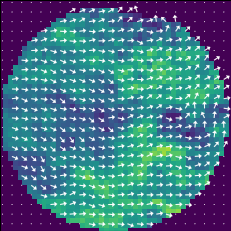}
            \caption*{CE-GAN}
        \end{subfigure}
        \begin{subfigure}[b]{0.27\textwidth}
            \includegraphics[width=0.99\textwidth]{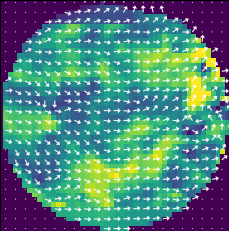}
            \caption*{GPOD-MF}
        \end{subfigure}
    \end{subfigure}
    \caption{Comparison of different model predictions for a single test snapshot at 10\% gaps. Gappy images formed by removing data outside of the red lines in the original image are fed into the models.}
    \label{fig:models10}
\end{figure}

\begin{figure}[htbp]
    \centering
    \begin{subfigure}[b]{0.95\textwidth}
    \centering
        \begin{subfigure}[b]{0.27\textwidth}
            \includegraphics[width=\textwidth]{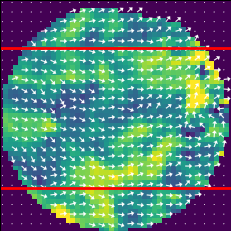}
            \caption*{Original}
        \end{subfigure}
        \begin{subfigure}[b]{0.27\textwidth}
            \includegraphics[width=\textwidth]{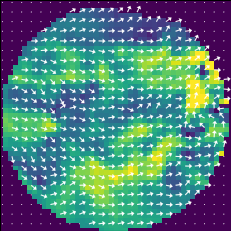}
            \caption*{UNet, MSE}
        \end{subfigure}
        \begin{subfigure}[b]{0.27\textwidth}
            \includegraphics[width=\textwidth]{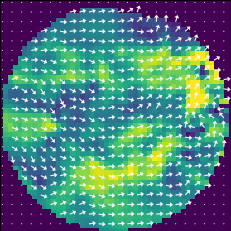}
            \caption*{UNet, huber}
        \end{subfigure}\\
        \begin{subfigure}[b]{0.27\textwidth}
            \includegraphics[width=\textwidth]{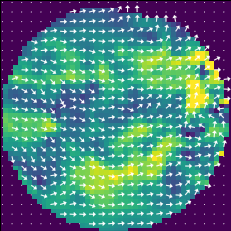}
            \caption*{UNETR}
        \end{subfigure}
        \begin{subfigure}[b]{0.27\textwidth}
            \includegraphics[width=\textwidth]{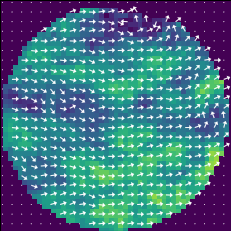}
            \caption*{CE-GAN}
        \end{subfigure}
        \begin{subfigure}[b]{0.27\textwidth}
            \includegraphics[width=\textwidth]{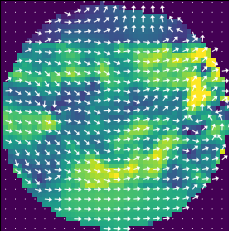}
            \caption*{GPOD-MF}
        \end{subfigure}
    \end{subfigure}
    \caption{Comparison of different model predictions for a single test snapshot at 25\% gaps. Gappy images formed by removing data outside of the red lines in the original image are fed into the models.}
    \label{fig:models25}
\end{figure}

Outputs from each model at an arbitrary snapshot are given for 10\% and 25\% gaps in Figures \ref{fig:models10} and \ref{fig:models25}.

\clearpage
\section*{NeurIPS Paper Checklist}

\begin{enumerate}

\item {\bf Claims}
    \item[] Question: Do the main claims made in the abstract and introduction accurately reflect the paper's contributions and scope?
    \item[] Answer: 
    \answerYes{} 
    \item[] Justification: We introduce our dataset, inpainting task, and present the benchmarking results. A literature review showing the lack of datasets available for the design of combustion machinery and lack of work done on realistic gap handling in fluids is given in Section \ref{sec:litrev}.

\item {\bf Limitations}
    \item[] Question: Does the paper discuss the limitations of the work performed by the authors?
    \item[] Answer: \answerYes{} 
    \item[] Justification: We describe the limitations of our work in Section \ref{sec:concl}.

\item {\bf Theory Assumptions and Proofs}
    \item[] Question: For each theoretical result, does the paper provide the full set of assumptions and a complete (and correct) proof?
    \item[] Answer: \answerNA{} 
    \item[] Justification: This is an empirical benchmarking study.

    \item {\bf Experimental Result Reproducibility}
    \item[] Question: Does the paper fully disclose all the information needed to reproduce the main experimental results of the paper to the extent that it affects the main claims and/or conclusions of the paper (regardless of whether the code and data are provided or not)?
    \item[] Answer: \answerYes{} 
    \item[] Justification: \\The code for training and testing the ML models is available at the github repo: \texttt{https://github.com/sambkr/EngineBench}.
    \\The data are available along with tutorials for engaging with the data at: \\ \texttt{https://www.kaggle.com/datasets/samueljbaker/enginebench-lsp-small/code}. \\In addition, full descriptions of the benchmark setup are given in Section \ref{sec:bench}.

\item {\bf Open access to data and code}
    \item[] Question: Does the paper provide open access to the data and code, with sufficient instructions to faithfully reproduce the main experimental results, as described in supplemental material?
    \item[] Answer: \answerYes{} 
    \item[] Justification: Links to the github repo, data, and notebook tutorials are all given at the webpage: \\
    \texttt{https://eng.ox.ac.uk/tpsrg/research/enginebench/}.

\item {\bf Experimental Setting/Details}
    \item[] Question: Does the paper specify all the training and test details (e.g., data splits, hyperparameters, how they were chosen, type of optimizer, etc.) necessary to understand the results?
    \item[] Answer: \answerYes{} 
    \item[] Justification: Training and testing details are given in Section \ref{sec:bench}, with further information on the train/test splits and loss functions given in Appendices \ref{app:perm} and \ref{app:loss} respectively.

\item {\bf Experiment Statistical Significance}
    \item[] Question: Does the paper report error bars suitably and correctly defined or other appropriate information about the statistical significance of the experiments?
    \item[] Answer: \answerYes{} 
    \item[] Justification: Error bars are reported in the main results table as well as in Figure \ref{fig:spectra}. These error bars were generated by training the models on different compositions of data in the training set as defined in Appendix \ref{app:perm} to ensure robustness.

\item {\bf Experiments Compute Resources}
    \item[] Question: For each experiment, does the paper provide sufficient information on the computer resources (type of compute workers, memory, time of execution) needed to reproduce the experiments?
    \item[] Answer: \answerYes{} 
    \item[] Justification: Training times for each model are given in Appendix \ref{app:modelsizes}. 
    
\item {\bf Code Of Ethics}
    \item[] Question: Does the research conducted in the paper conform, in every respect, with the NeurIPS Code of Ethics \url{https://neurips.cc/public/EthicsGuidelines}?
    \item[] Answer: \answerYes{} 
    \item[] Justification: The study involves a physics dataset, with no human participants or personally identifiable information. The results are reproducible with code and data that are well-documented in accordance with best practices. 

\item {\bf Broader Impacts}
    \item[] Question: Does the paper discuss both potential positive societal impacts and negative societal impacts of the work performed?
    \item[] Answer: \answerNA{} 
    \item[] Justification: This work is focussed on the study of turbulent airflows, with no human subjects.

\item {\bf Safeguards}
    \item[] Question: Does the paper describe safeguards that have been put in place for responsible release of data or models that have a high risk for misuse (e.g., pretrained language models, image generators, or scraped datasets)?
    \item[] Answer: \answerNA{} 
    \item[] Justification: The data are flow physics data, and the models consist of broadly distributed computer vision models.

\item {\bf Licenses for existing assets}
    \item[] Question: Are the creators or original owners of assets (e.g., code, data, models), used in the paper, properly credited and are the license and terms of use explicitly mentioned and properly respected?
    \item[] Answer: \answerYes{} 
    \item[] Justification: The citations required by the original publishers of the TCC-III data are given in Section~\ref{sec:data}, with the required acknowledgements to the original funding bodies given in the \textbf{Licensing} section. These requirements are also adhered to throughout our webpage, github repo and kaggle sites. 

\item {\bf New Assets}
    \item[] Question: Are new assets introduced in the paper well documented and is the documentation provided alongside the assets?
    \item[] Answer: \answerYes{} 
    \item[] Justification: We provide thorough documentation for interfacing with our code and data on the github repos and via tutorials on Kaggle. Our datasets include structured \texttt{http://schema.org} metadata that pass the rich results test \texttt{(https://search.google.com/test/rich-results}. 

\item {\bf Crowdsourcing and Research with Human Subjects}
    \item[] Question: For crowdsourcing experiments and research with human subjects, does the paper include the full text of instructions given to participants and screenshots, if applicable, as well as details about compensation (if any)? 
    \item[] Answer: \answerNA{} 
    \item[] Justification: This paper does not involve crowdsourcing nor research with human subjects.

\item {\bf Institutional Review Board (IRB) Approvals or Equivalent for Research with Human Subjects}
    \item[] Question: Does the paper describe potential risks incurred by study participants, whether such risks were disclosed to the subjects, and whether Institutional Review Board (IRB) approvals (or an equivalent approval/review based on the requirements of your country or institution) were obtained?
    \item[] Answer: \answerNA{} 
    \item[] Justification: This paper does not involve crowdsourcing nor research with human subjects.

\end{enumerate}

\clearpage
\section{Dataset supplementary material}
\paragraph{Webpage}
The project webpage is hosted at \texttt{https://eng.ox.ac.uk/tpsrg/research/enginebench/}. 
\paragraph{Subset}
The \texttt{EngineBench\_LSP\_small} data subset used in this paper is available at \texttt{https://www.kaggle.com/datasets/samueljbaker/enginebench-lsp-small/data}.
\paragraph{Code} 
Code for training and evaluating models in this study are available in \texttt{https://github.com/sambkr/EngineBench}, with instructions provided in the README of the repository.
\paragraph{DOI}
A persistent DOI for the EngineBench database is available: 10.34740/kaggle/ds/5000332.
\paragraph{Metadata}
The Croissant metadata record is available at the Kaggle repo: \texttt{https://www.kaggle.com/datasets/samueljbaker/enginebench}.
\paragraph{Tutorials}
Tutorials are attached to each of the Kaggle URLs. These notebooks are also consolidated on the project webpage. 
\paragraph{Distribution and maintenance}
The datasets are released via Kaggle and can be accessed at the persistent DOI or via the project webpage. \texttt{EngineBench\_LSP\_small} is not intended to be updated for reproducibility purposes, however we plan to add to the full \texttt{EngineBench} database with data from additional geometries on an approximately annual basis. In both cases, issues will be monitored at \texttt{https://github.com/sambkr/EngineBench/issues}. 
\paragraph{Licensing}
All EngineBench data are licensed via CC BY-NC-SA 4.0.

\section{Author statement}
All data is generated by the present authors and licensed via CC BY-NC-SA 4.0. The present authors bear responsibility in case of violation of rights.

\definecolor{darkblue}{RGB}{46,25, 110}

\newcommand{\dssectionheader}[1]{%
   \noindent\framebox[\columnwidth]{%
      {\fontfamily{phv}\selectfont \textbf{\textcolor{darkblue}{#1}}}
   }
}

\newcommand{\dsquestion}[1]{%
    {\noindent \fontfamily{phv}\selectfont \textcolor{darkblue}{\textbf{#1}}}
}

\newcommand{\dsquestionex}[2]{%
    {\noindent \fontfamily{phv}\selectfont \textcolor{darkblue}{\textbf{#1} #2}}
}

\newcommand{\dsanswer}[1]{%
   {\noindent #1 \medskip}
}

\section{Datasheet for datasets}

\dssectionheader{Motivation}

\dsquestionex{For what purpose was the dataset created?}{Was there a specific task in mind? Was there a specific gap that needed to be filled? Please provide a description.}

\dsanswer{EngineBench has been introduced to address the lack of open-source datasets that can be used to train ML models for propulsion system design. Existing datasets consist of synthetic data or do not accurately reflect the complex geometries and operating behaviours associated with propulsion systems. This dataset and the associated benchmark therefore lower the barrier to entry into ML for propulsion system designers.
}

\dsquestion{Who created this dataset (e.g., which team, research group) and on behalf of which entity (e.g., company, institution, organization)?}

\dsanswer{EngineBench has been created by the Thermal Propulsion Systems Research Group (TPSRG) at the University of Oxford. Note that the data for the TCC-III engine were originally gathered by the General Motors University of
Michigan Automotive Cooperative Research Laboratory, Engine Systems Division.
}

\dsquestionex{Who funded the creation of the dataset?}{If there is an associated grant, please provide the name of the grantor and the grant name and number.}

\dsanswer{This research was funded in whole or in part by the Engineering and Physical Sciences Research Council Prosperity Partnership, Grant No. EP/T005327/1.
}

\dsquestion{Any other comments?}

\bigskip
\dssectionheader{Composition}

\dsquestionex{What do the instances that comprise the dataset represent (e.g., documents, photos, people, countries)?}{ Are there multiple types of instances (e.g., movies, users, and ratings; people and interactions between them; nodes and edges)? Please provide a description.}

\dsanswer{Each instance consists of 2D particle image velocimetry (PIV) data, which represents the turbulent airflow patterns along a PIV plane in the TCC-III cylinder. The different instances are PIV images taken from different phase angles, engine cycles, and engine operating conditions.
}

\dsquestion{How many instances are there in total (of each type, if appropriate)?}

\dsanswer{There are 172393 instances on the lower swirl plane, 70125 on the tumble plane, and 33370 instances on the cross-tumble plane. The total number of instances is therefore 275888. 
}

\dsquestionex{Does the dataset contain all possible instances or is it a sample (not necessarily random) of instances from a larger set?}{ If the dataset is a sample, then what is the larger set? Is the sample representative of the larger set (e.g., geographic coverage)? If so, please describe how this representativeness was validated/verified. If it is not representative of the larger set, please describe why not (e.g., to cover a more diverse range of instances, because instances were withheld or unavailable).}

\dsanswer{EngineBench contains all the instances released in the TCC-III CFD input package.
}

\dsquestionex{What data does each instance consist of? “Raw” data (e.g., unprocessed text or images) or features?}{In either case, please provide a description.}

\dsanswer{Each instance consists of horizontal and vertical velocity data at each pixel (point in the PIV grid).
}

\dsquestionex{Is there a label or target associated with each instance?}{If so, please provide a description.}

\dsanswer{Each instance is associated with a specific phase angle, cycle number and operating condition, given by the vector encoding attributes included in the HDF5 files. 
}

\dsquestionex{Is any information missing from individual instances?}{If so, please provide a description, explaining why this information is missing (e.g., because it was unavailable). This does not include intentionally removed information, but might include, e.g., redacted text.}

\dsanswer{No.
}

\dsquestionex{Are relationships between individual instances made explicit (e.g., users’ movie ratings, social network links)?}{If so, please describe how these relationships are made explicit.}

\dsanswer{Yes, each instance is recorded from the same engine.
}

\dsquestionex{Are there recommended data splits (e.g., training, development/validation, testing)?}{If so, please provide a description of these splits, explaining the rationale behind them.}

\dsanswer{No. 
}

\dsquestionex{Are there any errors, sources of noise, or redundancies in the dataset?}{If so, please provide a description.}

\dsanswer{Yes, PIV as an experimental technique is affected by errors and noise. In the original experiments, Schiffmann \textit{et al.} \cite{schiffmann2016tcc} kept the overall average velocity error for the TCC-III PIV data to 1.5 to 8\% between the mid-intake and mid-compression strokes (90 to 270 CAD) depending on the PIV measurement plane. For more details, see \cite{schiffmann2016tcc}.
}

\dsquestionex{Is the dataset self-contained, or does it link to or otherwise rely on external resources (e.g., websites, tweets, other datasets)?}{If it links to or relies on external resources, a) are there guarantees that they will exist, and remain constant, over time; b) are there official archival versions of the complete dataset (i.e., including the external resources as they existed at the time the dataset was created); c) are there any restrictions (e.g., licenses, fees) associated with any of the external resources that might apply to a future user? Please provide descriptions of all external resources and any restrictions associated with them, as well as links or other access points, as appropriate.}

\dsanswer{EngineBench is self-contained.
}

\dsquestionex{Does the dataset contain data that might be considered confidential (e.g., data that is protected by legal privilege or by doctor-patient confidentiality, data that includes the content of individuals non-public communications)?}{If so, please provide a description.}

\dsanswer{No.
}

\dsquestionex{Does the dataset contain data that, if viewed directly, might be offensive, insulting, threatening, or might otherwise cause anxiety?}{If so, please describe why.}

\dsanswer{No.
}

\dsquestionex{Does the dataset relate to people?}{If not, you may skip the remaining questions in this section.}

\dsanswer{No.
}

\dsquestionex{Does the dataset identify any subpopulations (e.g., by age, gender)?}{If so, please describe how these subpopulations are identified and provide a description of their respective distributions within the dataset.}

\dsanswer{No.
}

\dsquestionex{Is it possible to identify individuals (i.e., one or more natural persons), either directly or indirectly (i.e., in combination with other data) from the dataset?}{If so, please describe how.}

\dsanswer{No.
}

\dsquestionex{Does the dataset contain data that might be considered sensitive in any way (e.g., data that reveals racial or ethnic origins, sexual orientations, religious beliefs, political opinions or union memberships, or locations; financial or health data; biometric or genetic data; forms of government identification, such as social security numbers; criminal history)?}{If so, please provide a description.}

\dsanswer{No.
}

\dsquestion{Any other comments?}

\bigskip
\dssectionheader{Collection Process}

\dsquestionex{How was the data associated with each instance acquired?}{Was the data directly observable (e.g., raw text, movie ratings), reported by subjects (e.g., survey responses), or indirectly inferred/derived from other data (e.g., part-of-speech tags, model-based guesses for age or language)? If data was reported by subjects or indirectly inferred/derived from other data, was the data validated/verified? If so, please describe how.}

\dsanswer{The PIV data were acquired by running motored (operation with pure air) experiments on the TCC-III engine.
}

\dsquestionex{What mechanisms or procedures were used to collect the data (e.g., hardware apparatus or sensor, manual human curation, software program, software API)?}{How were these mechanisms or procedures validated?}

\dsanswer{A PIV setup was used to acquire the data, described in detail along with the validation process in \cite{schiffmann2016tcc}.
}

\dsquestion{If the dataset is a sample from a larger set, what was the sampling strategy (e.g., deterministic, probabilistic with specific sampling probabilities)?}

\dsanswer{The dataset is not a sample.
}

\dsquestion{Who was involved in the data collection process (e.g., students, crowdworkers, contractors) and how were they compensated (e.g., how much were crowdworkers paid)?}

\dsanswer{The original data were collected by the General Motors University of Michigan Automotive Cooperative Research Laboratory, Engine Systems Division. EngineBench was created and processed by the authors of this work for no additional payment outside of typical salary and stipend.
}

\dsquestionex{Over what timeframe was the data collected? Does this timeframe match the creation timeframe of the data associated with the instances (e.g., recent crawl of old news articles)?}{If not, please describe the timeframe in which the data associated with the instances was created.}

\dsanswer{The original data were gathered across a six month experimental campaign in 2014.
}

\dsquestionex{Were any ethical review processes conducted (e.g., by an institutional review board)?}{If so, please provide a description of these review processes, including the outcomes, as well as a link or other access point to any supporting documentation.}

\dsanswer{No.
}

\dsquestionex{Does the dataset relate to people?}{If not, you may skip the remaining questions in this section.}

\dsanswer{No.
}

\dsquestion{Did you collect the data from the individuals in question directly, or obtain it via third parties or other sources (e.g., websites)?}

\dsanswer{
}

\dsquestionex{Were the individuals in question notified about the data collection?}{If so, please describe (or show with screenshots or other information) how notice was provided, and provide a link or other access point to, or otherwise reproduce, the exact language of the notification itself.}

\dsanswer{
}

\dsquestionex{Did the individuals in question consent to the collection and use of their data?}{If so, please describe (or show with screenshots or other information) how consent was requested and provided, and provide a link or other access point to, or otherwise reproduce, the exact language to which the individuals consented.}

\dsanswer{
}

\dsquestionex{If consent was obtained, were the consenting individuals provided with a mechanism to revoke their consent in the future or for certain uses?}{If so, please provide a description, as well as a link or other access point to the mechanism (if appropriate).}

\dsanswer{
}

\dsquestionex{Has an analysis of the potential impact of the dataset and its use on data subjects (e.g., a data protection impact analysis) been conducted?}{If so, please provide a description of this analysis, including the outcomes, as well as a link or other access point to any supporting documentation.}

\dsanswer{
}

\dsquestion{Any other comments?}

\dsanswer{
}

\bigskip
\dssectionheader{Preprocessing/cleaning/labeling}

\dsquestionex{Was any preprocessing/cleaning/labeling of the data done (e.g., discretization or bucketing, tokenization, part-of-speech tagging, SIFT feature extraction, removal of instances, processing of missing values)?}{If so, please provide a description. If not, you may skip the remainder of the questions in this section.}

\dsanswer{A typical PIV post-processing routine was conducted to map the raw images to vector fields, fully detailed in \cite{schiffmann2016tcc}.
}

\dsquestionex{Was the “raw” data saved in addition to the preprocessed/cleaned/labeled data (e.g., to support unanticipated future uses)?}{If so, please provide a link or other access point to the “raw” data.}

\dsanswer{No. If this is required, the original collectors of the TCC-III data should be contacted.
}

\dsquestionex{Is the software used to preprocess/clean/label the instances available?}{If so, please provide a link or other access point.}

\dsanswer{All vector fields were originally calculated using a commercial PIV code (DaVis 8.x, LaVision). Python was used to further process the data for use in EngineBench.
}

\dsquestion{Any other comments?}

\dsanswer{
}

\bigskip
\dssectionheader{Uses}

\dsquestionex{Has the dataset been used for any tasks already?}{If so, please provide a description.}

\dsanswer{Yes, EngineBench has been used to establish an inpainting benchmark as described in this paper. 
}

\dsquestionex{Is there a repository that links to any or all papers or systems that use the dataset?}{If so, please provide a link or other access point.}

\dsanswer{No.
}

\dsquestion{What (other) tasks could the dataset be used for?}

\dsanswer{EngineBench can be used to train and benchmark ML models across a range of standard tasks in turbulence data processing, such as super-resolution, 2D to 3D reconstruction, developing closure models or solvinng inverse problems. 
}

\dsquestionex{Is there anything about the composition of the dataset or the way it was collected and preprocessed/cleaned/labeled that might impact future uses?}{For example, is there anything that a future user might need to know to avoid uses that could result in unfair treatment of individuals or groups (e.g., stereotyping, quality of service issues) or other undesirable harms (e.g., financial harms, legal risks) If so, please provide a description. Is there anything a future user could do to mitigate these undesirable harms?}

\dsanswer{No.
}

\dsquestionex{Are there tasks for which the dataset should not be used?}{If so, please provide a description.}

\dsanswer{The dataset is suitable for use in a range of industrial flow physics scenarios.
}

\dsquestion{Any other comments?}

\bigskip
\dssectionheader{Distribution}

\dsquestionex{Will the dataset be distributed to third parties outside of the entity (e.g., company, institution, organization) on behalf of which the dataset was created?}{If so, please provide a description.}

\dsanswer{Yes, EngineBench is distributed via Kaggle.
}

\dsquestionex{How will the dataset will be distributed (e.g., tarball on website, API, GitHub)}{Does the dataset have a digital object identifier (DOI)?}

\dsanswer{Via Kaggle, with the DOI: 10.34740/kaggle/ds/5000332.

}

\dsquestion{When will the dataset be distributed?}

\dsanswer{EngineBench is currently available.
}

\dsquestionex{Will the dataset be distributed under a copyright or other intellectual property (IP) license, and/or under applicable terms of use (ToU)?}{If so, please describe this license and/or ToU, and provide a link or other access point to, or otherwise reproduce, any relevant licensing terms or ToU, as well as any fees associated with these restrictions.}

\dsanswer{EngineBench is licensed via CC BY-SA NC 4.0.
}

\dsquestionex{Have any third parties imposed IP-based or other restrictions on the data associated with the instances?}{If so, please describe these restrictions, and provide a link or other access point to, or otherwise reproduce, any relevant licensing terms, as well as any fees associated with these restrictions.}

\dsanswer{The original collectors of the TCC-III data ask that any published use of the TCC engine simulation geometry and/or data be acknowledged with the following statement:\\
“The TCC engine work has been funded by General Motors through the General Motors University of Michigan Automotive Cooperative Research Laboratory, Engine Systems Division.”
}

\dsquestionex{Do any export controls or other regulatory restrictions apply to the dataset or to individual instances?}{If so, please describe these restrictions, and provide a link or other access point to, or otherwise reproduce, any supporting documentation.}
no.
\dsanswer{
}

\dsquestion{Any other comments?}

\dsanswer{
}

\bigskip
\dssectionheader{Maintenance}

\dsquestion{Who will be supporting/hosting/maintaining the dataset?}

\dsanswer{The Thermal Propulsion Systems Research Group (TPSRG) at the University of Oxford \texttt{https://eng.ox.ac.uk/tpsrg/about-us/}.
}

\dsquestion{How can the owner/curator/manager of the dataset be contacted (e.g., email address)?}

\dsanswer{Via GitHub \texttt{https://github.com/sambkr/EngineBench/issues} or any of the author email addresses.
}

\dsquestionex{Is there an erratum?}{If so, please provide a link or other access point.}

\dsanswer{Yes, via \texttt{https://github.com/sambkr/EngineBench/issues}.
}

\dsquestionex{Will the dataset be updated (e.g., to correct labeling errors, add new instances, delete instances)?}{If so, please describe how often, by whom, and how updates will be communicated to users (e.g., mailing list, GitHub)?}

\dsanswer{Yes, errors will be monitored on the GitHub issues page and rectified accordingly. Additional data pertaining to other engine geometries are intended to be added to EngineBench on an approximately annual basis.
}

\dsquestionex{If the dataset relates to people, are there applicable limits on the retention of the data associated with the instances (e.g., were individuals in question told that their data would be retained for a fixed period of time and then deleted)?}{If so, please describe these limits and explain how they will be enforced.}

\dsanswer{N/A.
}

\dsquestionex{Will older versions of the dataset continue to be supported/hosted/maintained?}{If so, please describe how. If not, please describe how its obsolescence will be communicated to users.}

\dsanswer{Yes, version histories and logs from Kaggle repositories will be used to inform users of changes. Changes will also be communicated to users via the landing page \texttt{https://eng.ox.ac.uk/tpsrg/research/enginebench/}.
}

\dsquestionex{If others want to extend/augment/build on/contribute to the dataset, is there a mechanism for them to do so?}{If so, please provide a description. Will these contributions be validated/verified? If so, please describe how. If not, why not? Is there a process for communicating/distributing these contributions to other users? If so, please provide a description.}

\dsanswer{Yes, for additions to the EngineBench database please contact one of the members of the TPSRG \texttt{https://eng.ox.ac.uk/tpsrg/about-us/academics/}.
}

\dsquestion{Any other comments?}

\dsanswer{
}

\end{document}